\documentclass[twocolumn,showpacs,amssymb,10pt]{revtex4}
\usepackage{amssymb}
\usepackage{amsmath}
\usepackage{subfigure}
\usepackage{natbib}
\usepackage{amsfonts}
\usepackage{graphicx}
\usepackage{dcolumn}
\usepackage{bm}
\usepackage{epsfig}
\usepackage{float}
\usepackage{makeidx}
\usepackage{color}
\usepackage{xcolor}
\usepackage{cases}
\graphicspath{{figs/}}
\usepackage{longtable}
\usepackage{mathtools}
\usepackage{multirow}
\usepackage{diagbox}
\usepackage{makecell}

\begin{document}

\preprint{}
%
%
\title{Resummation of Divergent Series and Quantum Phase Transitions in Kitaev Chains with Long-Range Hopping}
%
%

\author{Hao Fu$^{1}$ and Peiqing Tong$^{2,\rm \dag}$\footnotetext{$^{\rm \dag}$Electronic mail: pqtong@njnu.edu.cn}}
\address{$^{1}$Department of Mathematics and Physics, Bengbu University, Bengbu, 233030, People's Republic of China\\
$^{2}$Ministy of Education Key Laboratory of NSLSCS and School of Physics and Technology, Nanjing Normal University, Nanjing, 210023, People's Republic of China}

\date{\today \\[0.3cm]}

%
%
\begin{abstract}
  We study the quantum phase transitions (QPTs) in extended Kitaev chains with long-range ($1/r^{\alpha}$) hopping. It has been known that there are two QPT points at $\mu=\mu_0(\alpha)$ and $\mu_\pi(\alpha)$ ($\mu$ the chemical potential) corresponding respectively to
  the summation of $\sum_{m=1}^{\infty}m^{-\alpha}$ and $\sum_{m=1}^{\infty}(-1)^{m-1}m^{-\alpha}$, for $\alpha>1$. When $\alpha\le0$, the two series divergent and it is practically assumed that no QPTs exist, as nothing meaningful can be concluded from divergent series. However, we find that there appear to exist two second order QPTs at $\mu=\mu_0(0)$ and $\mu_\pi(0)$ for $\alpha=0$, and one second order QPT at $\mu=\mu_\pi(\alpha)$ for $-2<\alpha<0$. The $\mu_0(0)$ and $\mu_\pi(-2<\alpha\leq0)$
   correspond to the resummation of the divergent series obtained by analytic continuation of the Riemann $\zeta$ function and Dirichlet $\eta$ function, respectively.
   For $\alpha\le-2$, there is indeed no QPT point in the system.
   Moreover, it is found that the quasiparticle energy spectra are discontinuous functions of the wave vector $k$ and have two branches for $\alpha<0$. This is quite different from that in the case of $\alpha>0$ and induces topological phases with the winding number $\omega:=\pm1/2$.
  At the same time, the von Neumann entropy follows power law of the length $L$ of subchains both in gapped and non-gapped regions.
  For comparison, we study the QPTs, topological properties, and von Neumann entropy of the systems with $\alpha>0$ for this long-range hopping chain.
\end{abstract}
\maketitle

%
%
Divergent series arise quite frequently in physics, such as in perturbation expansion of quantum field theory and in partition function calculation in statistical mechanics.
Therefore, how to deal with divergent series is a very important problem in physics.
Mathematicians from as early as Euler's time have motivated various types of formal resummation of divergent series, such as
the Abel summation formula, Ces${\rm\grave{a}}$ro summation, and Borel summation method \cite{Apo,Har,Edw,Tit2},
and obtain some counterintuitive and seemingly absurd results, such as $\sum_{m=1}^{\infty} m:=-\frac{1}{12}$ \cite{Sup5}.
In fact, in 1948, Casimir predicted what is now called Casimir force, by applying resummation to the divergent series of the self-energy of electromagnetic field between two plates (corresponding to $\sum_{m=1}^{\infty} m^{3}:=\frac{1}{120}$) \cite{Cas,Eli}. His prediction of the attractive interaction between two plates in the electromagnetic field has been verified and measured in laboratory \cite{Lam,Moh,Muk,Bos}. This is the first known explicit demonstration how resummation can be applied to a physical discovery.
While there are some applications of the resummation of divergent series in quantum field theory, gravitation and cosmology \cite{Eli2}, it is still an essential open and interesting question to find the application of the resummation of divergent series in other physical systems, especially in condensed matter systems.




This letter discusses an application of resummation of divergent series to an extended Kitaev chains with long-range (LR) interactions. Systems with LR interactions such as wave-particle interactions, gravitational forces, and Coulomb forces, are rather common in nature and laboratory \cite{Dau,Cam,Def2,Cam2,Lev}.
It is well known that in such systems, various physical properties are different from those in systems with short-range (SR) interactions.
Example differences include convexity of free energy \cite{Ell,Bou}, inequivalence of various statistical mechanical ensembles \cite{Coh},
conformal symmetry breaking\cite{Vod,Vod2,Are}, anomalous dynamical criticality \cite{Def,Dut}, massive Majorana edge modes \cite{Vod2,Pat,Jag,Ren,Viy}, half-integer winding number \cite{Viy,Pez}, failure of the Uhlmann phase approach \cite{Bha}, anomalous Lieb-Robinson bound \cite{Her}, significant reduction of the transition rate of bound states \cite{Rad}, and breakdown or protection of quasilocality \cite{Eis2,Cev}, {\sl etc}.
Many of these properties are confirmed experimentally by using Floquet engineering via an applied external AC field \cite{Ben,LiL2}, neutral atoms or trapped ions loaded onto an optical lattice coupled to photonic modes \cite{Sch,Dou,Bri,Fri,Kim,Sch2,Gop,Per,Ber,Nev,Mol}, planar Josephson junctions in proximity to a 2D electron gas \cite{Liu2,Liu}, or Shiba bound states induced in a chain of magnetic impurities on top of an $s$-wave superconductor \cite{Pie,Pie2}, {\sl etc}.

\begin{figure*}[tb]
 \centering
\includegraphics[width=.9\linewidth]{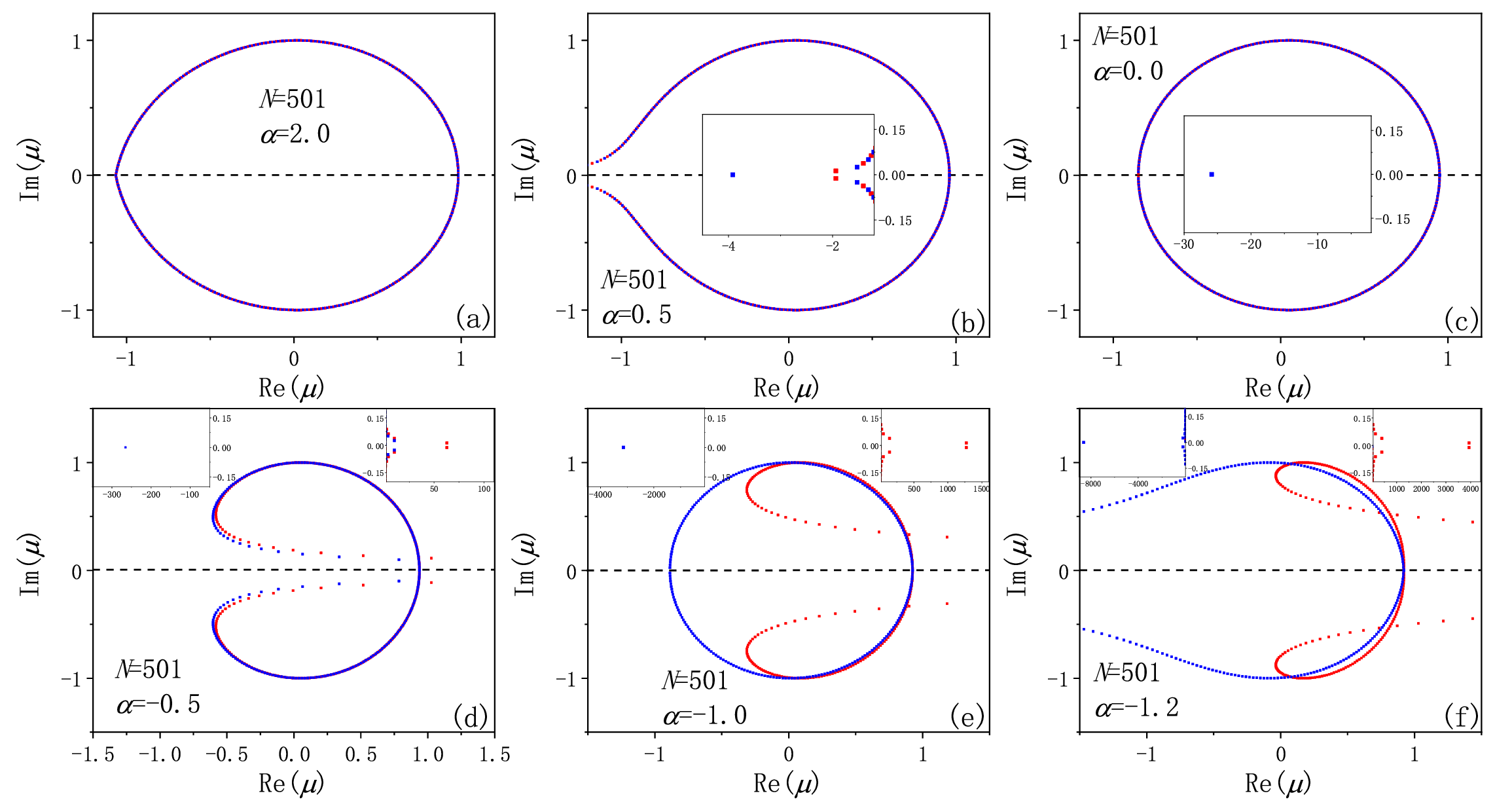}
 \caption{\label{1} (Color online) The Lee-Yang zeros of $\varepsilon_k=0$ in the complex $\mu$ plane.
 (a)-(f) $\alpha=2$, $0.5$, $0$, $-0.5$, $-1$, and $-1.2$, respectively.
 The parameter $f=0.1$ and the lattice number $N=501$.
 The blue (red) points correspond to $k=\pm2n\cdot2\pi/N$ ($\pm(2n-1)\cdot2\pi/N$) with $n=1,2,3,\dots$.
 The insets of (b)-(c) show the zeros with large negative real parts.
 The left (right) upper insets of (d), (e), and (f) show the zeros with large negative (positive) real parts of the zeros.
 The black dashed line corresponds to the real axis.
 }
\end{figure*}

While most works focus on the systems with power-decay LR interaction as $1/r^\alpha$ ($\alpha>0$), the systems with power-growth LR interaction, {\sl i.e.} $\alpha<0$, are rarely discussed \cite{Sch4,Col,Dav,Aga}.
In this letter, we first extend Kitaev chain with LR hopping and study its physical behavior for $\alpha\le0$, which leads to some divergent series.
Our numerical calculation of the Lee-Yang zeros indicate the existence of the QPTs which correspond to the results of resumming those divergent series.
We then study various properties, such as topological winding numbers and von Neumann entropy, of the system related to these QPTs.

We consider the Kitaev chain with LR hopping whose Hamiltonian is \cite{Sup}
\begin{equation}\label{ham}
\begin{split}
  H=&-\sum_{n=1}^{N}[tc_n^\dagger c_{n+1}+\Delta c_n^\dagger c_{n+1}^\dagger+
  \mu(c_n^\dagger c_{n}-\frac{1}{2})]\\
  &-\sum_{n=1}^{N}\sum_{m=2}^{[N/2]}\frac{f}{m^\alpha}c_n^\dagger c_{n+m}+{\rm H.C.},
\end{split}
\end{equation}
where $c_n^\dagger$ ($c_n$) is a fermionic creation (annihilation) operator at site $n$. $t$, $\Delta$, and $f$ are the strengths of SR hopping, SR pairing, and LR hopping, respectively. $\mu$ is the chemical potential.
In this letter, we only study the case of odd $N$.
We set $t=\Delta=1$ without losing generality and take the periodic boundary conditions $c_{n+N}=c_{n}$.
The Hamiltonian (\ref{ham}) can be diagonalized as $H=\sum_k\varepsilon_k(2\gamma_k^\dagger\gamma_k-1)$ \cite{Sup,Sac,Lieb} with quasiparticle energy spectrum (QES)
\begin{equation}
  \varepsilon_k=\sqrt{[g_k(\alpha)+\mu]^2+\sin^2 k},
  \label{eq:qes}
\end{equation}
where
\begin{equation}
\begin{split}
  g_k(\alpha)=\cos k+\sum_{m=2}^{[N/2]}\frac{f}{m^\alpha}\cos(mk),
\end{split}
\end{equation}
with $k=\frac{2\pi h}{N}$ and $h$ is an integer from $[-N/2]$ to $[N/2]$.

In order to find the QPTs in the chain, we study the distribution of the Lee-Yang zeros $\varepsilon_k=0$ in the complex $\mu$ plane \cite{Y-L,L-Y,Ton} which yields
\begin{equation}\label{lyz}
\begin{split}
  \mu_k(\alpha)=-g_k(\alpha)+i\cdot\sin k.
\end{split}
\end{equation}
When $k=0$ and $\pi$, $\mu_k(\alpha)$ are real numbers which correspond to the QPT points, {\sl i.e}.
\begin{subequations}
    \begin{align}
    \mu_0(\alpha)=-1-\sum_{m=2}^{[N/2]}\frac{f}{m^\alpha}=-(1-f)-f\varphi(\alpha,N)\label{m0} \\
    \shortintertext{and}
    \mu_\pi(\alpha)=1-\sum_{m=2}^{[N/2]}\frac{(-1)^{m}f}{m^\alpha}=(1-f)+f\psi(\alpha,N),\label{mp}
    \end{align}
\end{subequations}
where $\varphi(\alpha,N)\equiv\sum_{m=1}^{[N/2]}m^{-\alpha}$ and $\psi(\alpha,N)\equiv\sum_{m=1}^{[N/2]}(-1)^{m-1}m^{-\alpha}$, respectively.
In the thermodynamic limit, for $\alpha>1$, $\varphi(\alpha,\infty)$ is convergent, which is the famous Riemann zeta function $\zeta(\alpha)$ \cite{Edw,Tit2,Sup}.
For $\alpha>0$, $\psi(\alpha,\infty)$ is convergent, which is the Dirichlet eta function $\eta(\alpha)$ \cite{Sup,Mil}.
Note that both $\varphi(\alpha,\infty)$ for $\alpha\le1$ and $\psi(\alpha,\infty)$ for $\alpha\le0$ are divergent.

\begin{figure}
 \centering
\includegraphics[width=1.0\linewidth]{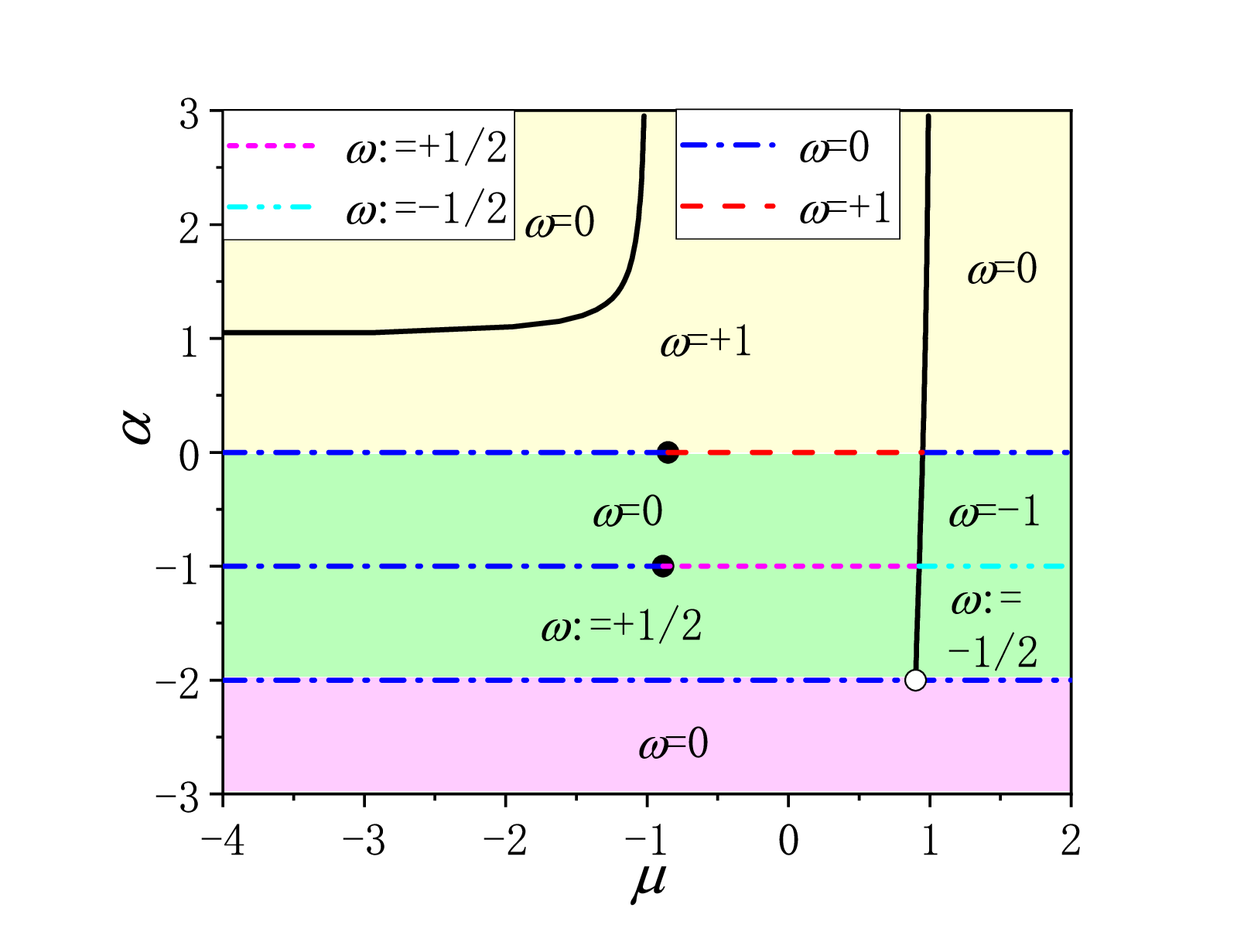}
 \caption{\label{ph} (Color online) The phase diagram of the LR Kitaev chain.
 The two black solid lines and the two black dot lines represent the critical region of the chain.
 The entropy increases logarithmically in the white plane and exponentially in the yellow plane with increasing $L$.
 }
\end{figure}

For finite $N$, we can obtain the Lee-Yang zeros in the complex $\mu$ plane numerically. According to the intersection of zeros and the real axis or the zeros converging to the real axis under $N\rightarrow\infty$, we can obtain the QPT points.
Fig. \ref{1} shows six typical cases of zeros with finite $N$ and different $\alpha$.
On these grounds, the phase diagram is shown roughly\emph{\emph{}} in Fig. \ref{ph}.

In what follows, we will give detailed discussions of the QPTs and their relation with the resummation of divergent series for different $\alpha$. In additions, we also discuss that the topological properties and entanglement in the phases and on the QPT points.

\emph{The case of $\alpha>0$}.---
When $\alpha>1$, both the $\varphi(\alpha,\infty)$ and $\psi(\alpha,\infty)$ are convergent.
Therefore, there are two QPT points at $\mu=\mu_0(\alpha)$ and $\mu_\pi(\alpha)$ obtained by Eqs. ({\ref{m0}})-({\ref{mp}}).
It can be conformed by studying the distribution of Lee-Yang zeros.
In Fig. \ref{1}(a), we show a typical example of $\alpha=2$.
It is obvious that the zeros converge to the real axis at $\mu=\mu_0(\alpha)$ and $\mu_\pi(\alpha)$ as $N\rightarrow\infty$.
Furthermore, it can be seen that $\mu_0\ne-\mu_\pi$.
It means that QPT points are no longer symmetric across the line $\mu=0$.
It is due to symmetry breaking under the transformation $c_n=(-1)^nc_{n}^\dagger$ by the LR hopping \cite{Vod}.

When $0<\alpha\le1$, $\psi(\alpha,\infty)$ is also convergent.
Similarly, we can obtain the $\mu_{\pi}(\alpha)$ from Eq. ({\ref{mp}}).
We also find that the zeros tend to the real axis at $\mu=\mu_{\pi}(\alpha)$ as $N\rightarrow\infty$ [see Fig. \ref{1}(b)].
However, the $\varphi(\alpha,\infty)$ and the corresponding $\mu_0(\alpha)$ are divergent.
Correspondingly,
the real parts of zeros corresponding to $k=0$ and $k\rightarrow0$ tend to $-\infty$ for $N\rightarrow\infty$ [see Fig. \ref{1}(b)].
It means that only one QPT point exists at $\mu=\mu_\pi(\alpha)$.

To address the properties of QPTs, we study the ground-state energy density $e=-\frac{1}{N}\sum_{k}\varepsilon_{k}$ as a function of $\mu$, and find that the second order partial derivative has singularity at $\mu=\mu_0(\alpha)$ and $\mu_\pi(\alpha)$ \cite{Sup}, meaning that the QPTs are second order.
Furthermore, we study the topological properties of the system \cite{Sup} and find that the phases for $\mu<\mu_0(\alpha)$ or $\mu>\mu_\pi(\alpha)$ are topologically trivial.
In the region $\mu_0(\alpha)<\mu<\mu_\pi(\alpha)$ ($\alpha>1$) and $\mu<\mu_\pi(\alpha)$ ($0<\alpha\le1$), the phase is topologically nontrivial with winding number $\omega=+1$ [see Fig. \ref{ph}].
Finally, it is found that the von Neumann entropy $S(L)\sim(c_{\rm eff}/3)\log L$, where $L$ is the length of the subchain and $c_{\rm eff}$ \cite{Vod} the effective central charge.
The results are briefly summarised as follows (see the details in the Supplemental Material \cite{Sup}):
($\rm{\romannumeral1}$) when $\alpha>0$ the effective central charge $c_{\rm eff}=0$ in the gapped region (off the critical lines).
($\rm{\romannumeral2}$) On the critical line $\mu=\mu_{\pi}(\alpha>0)$, $c_{\rm eff}=1/2$.
($\rm{\romannumeral3}$) On the critical line $\mu=\mu_{0}(\alpha)$, for $\alpha\ge2$, the $c_{\rm eff}$ equal to 1/2 and the quasiparticle velocity $\varepsilon'_{k\rightarrow0}(\alpha)$ is finite.
Whereas for $1<\alpha<2$, $c_{\rm eff}$ gradually decreases from $1/2$ to 0 as $\alpha$ decreasing and the quasiparticle velocity $\varepsilon'_{k\rightarrow0}(\alpha)\sim k^{\alpha-2}$ \cite{Sup}.
These results are quite different from those of the Kitaev chains with LR pairing \cite{Vod}.

\begin{figure}
 \centering
\includegraphics[width=.9\linewidth]{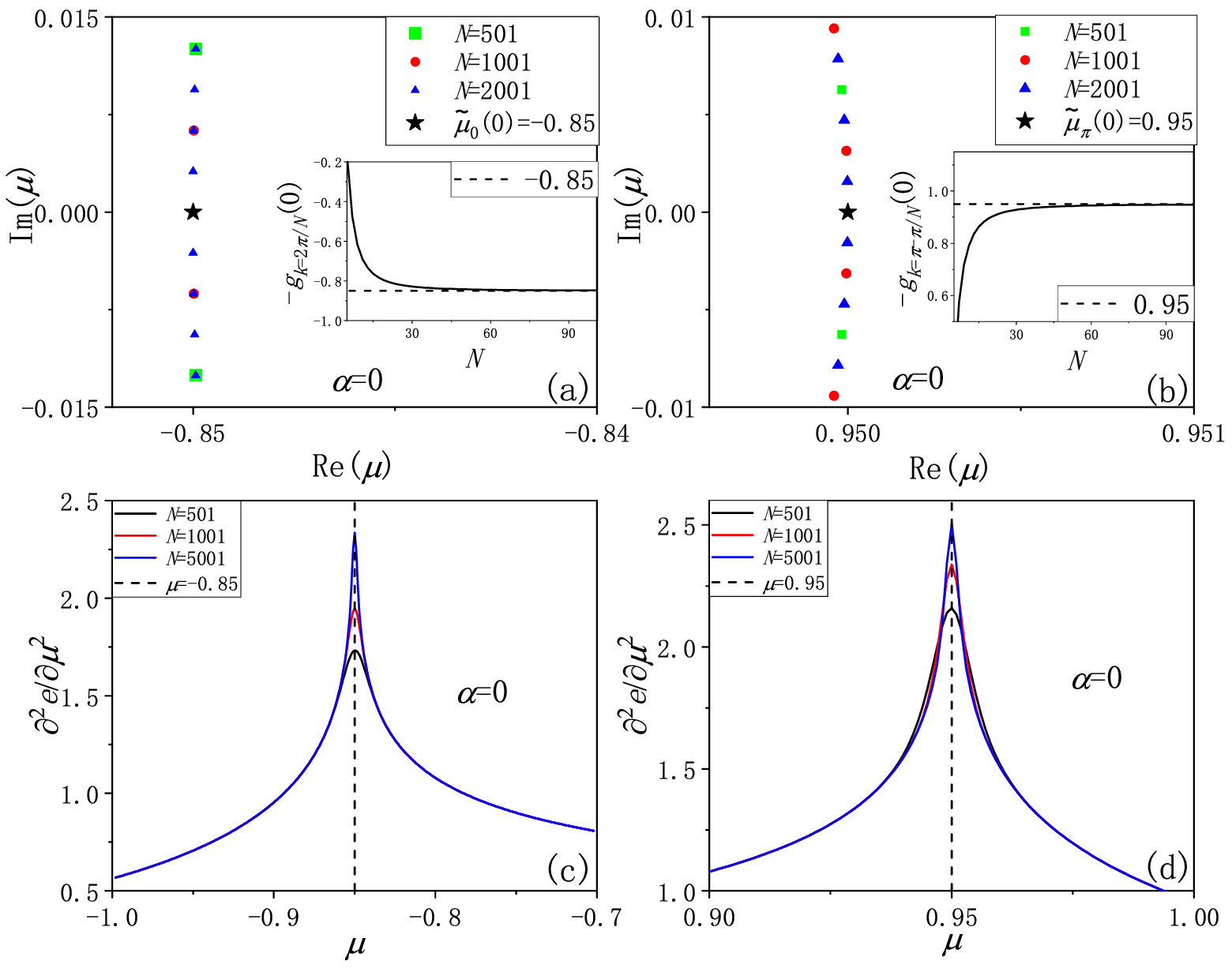}
 \caption{\label{3} (Color online) (a)-(b) The zeros are distributed near the QPT points.
 The two insets show the real parts of zeros $-g_{k=2\pi/N}(0)$ and $-g_{k=\pi-\pi/N}(0)$ as functions of $N$, respectively.
 (c)-(d) The second order partial derivative of the ground-state energy density as functions of $\mu$.
 ($\alpha=0$)
 }
\end{figure}

\emph{The case of $\alpha=0$}.---
Fig. \ref{1}(c) shows the zeros in the complex $\mu$ plane for $\alpha=0$.
Similar to that in the case of $0<\alpha<1$, the zero of $k=0$ tends to $-\infty$.
However, the zeros converge to real axis at $\mu=\tilde{\mu}_0(0)$ for $k\rightarrow0$.
This is different from that in the case of $0<\alpha<1$.
To see the convergence of the zeros more clearly, the zeros for different $N$ are shown in Fig. \ref{3}(a).
It is obvious that the zeros closest to the real axis [corresponding to $k=\pm(2\pi/N)$] tend to the real axis with increasing $N$ (see the inset).
For $k\rightarrow\pi$, the situation is similar. The zeros converge to real axis at $\mu=\tilde{\mu}_\pi(0)$ [see Fig. \ref{3}(b)].
It means that there are two QPT points at $\mu=\tilde{\mu}_{0}(0)$ and $\tilde{\mu}_{\pi}(0)$ in the thermodynamic limit.
This is different from that of the Kitaev chain with LR pairing in which only one QPT point exists for $\alpha=0$ \cite{Vod}.

On the other hand, for $\alpha=0$, both the $\varphi(0,\infty)$ and $\psi(0,\infty)$ are divergent.
The values of $\varphi(0,\infty)$ and $\psi(0,\infty)$ cannot be found from the series $\sum_{m=1}^{\infty}m^{-\alpha}$ and $\sum_{m=1}^{\infty}(-1)^{m-1}m^{-\alpha}$ directly.
However, the Riemann zeta function $\zeta(\alpha=0)$ and Dirichlet eta function $\eta(\alpha=0)$ can be obtained by analytic continuation \cite{Sup,Edw,Tit2}.
Then, $\varphi(0,\infty)$ and $\psi(0,\infty)$ can be defined \cite{Sup} as
\begin{subequations}
    \begin{align}
    \varphi(0,\infty)=\sum_{m=1}^{\infty}m^0:=\zeta(0)=-1/2\label{z0} \\
    \shortintertext{and}
    \psi(0,\infty)=\sum_{m=1}^{\infty}(-1)^{m-1}:=\eta(0)=1/2.\label{e0}
    \end{align}
\end{subequations}
From Eqs. (\ref{m0})-(\ref{mp}) and (\ref{z0})-(\ref{e0}), we can obtain a finite $\mu_0(0):=-0.85$ and $\mu_\pi(0):=0.95$ for $f=0.1$.
Surprisingly, they are the same as the values $\tilde{\mu}_0(0)$ and $\tilde{\mu}_\pi(0)$ obtained by Lee-Yang zeros numerically.
Therefore, the two QPT points at $\alpha=0$ correspond to the resummation of divergent series.

Similarly, we study the second order partial derivative of the ground-state energy density.
The results are shown in Fig. \ref{3}(c)-(d) near $\mu_0(0)$ and $\mu_\pi(0)$, respectively.
It is clear that the QPTs are second order.
Moreover, the winding numbers $\omega=0$ for $\mu<\mu_0(\alpha)$ or $\mu>\mu_\pi(\alpha)$ and $\omega=+1$ for $\mu_0(\alpha)<\mu<\mu_\pi(\alpha)$ \cite{Sup}. They are similar to those in the case of $\alpha\ge2$.
Furthermore, the von Neumann entropy $S(L)\sim(c_{\rm eff}/3)\log L$ at $\mu=\mu_0(0)$ and $\mu_\pi(0)$ with $c_{\rm eff}=1/2$.
And the $c_{\rm eff}=0$ in the gapped region \cite{Sup}.

\begin{figure}
 \centering
\includegraphics[width=.9\linewidth]{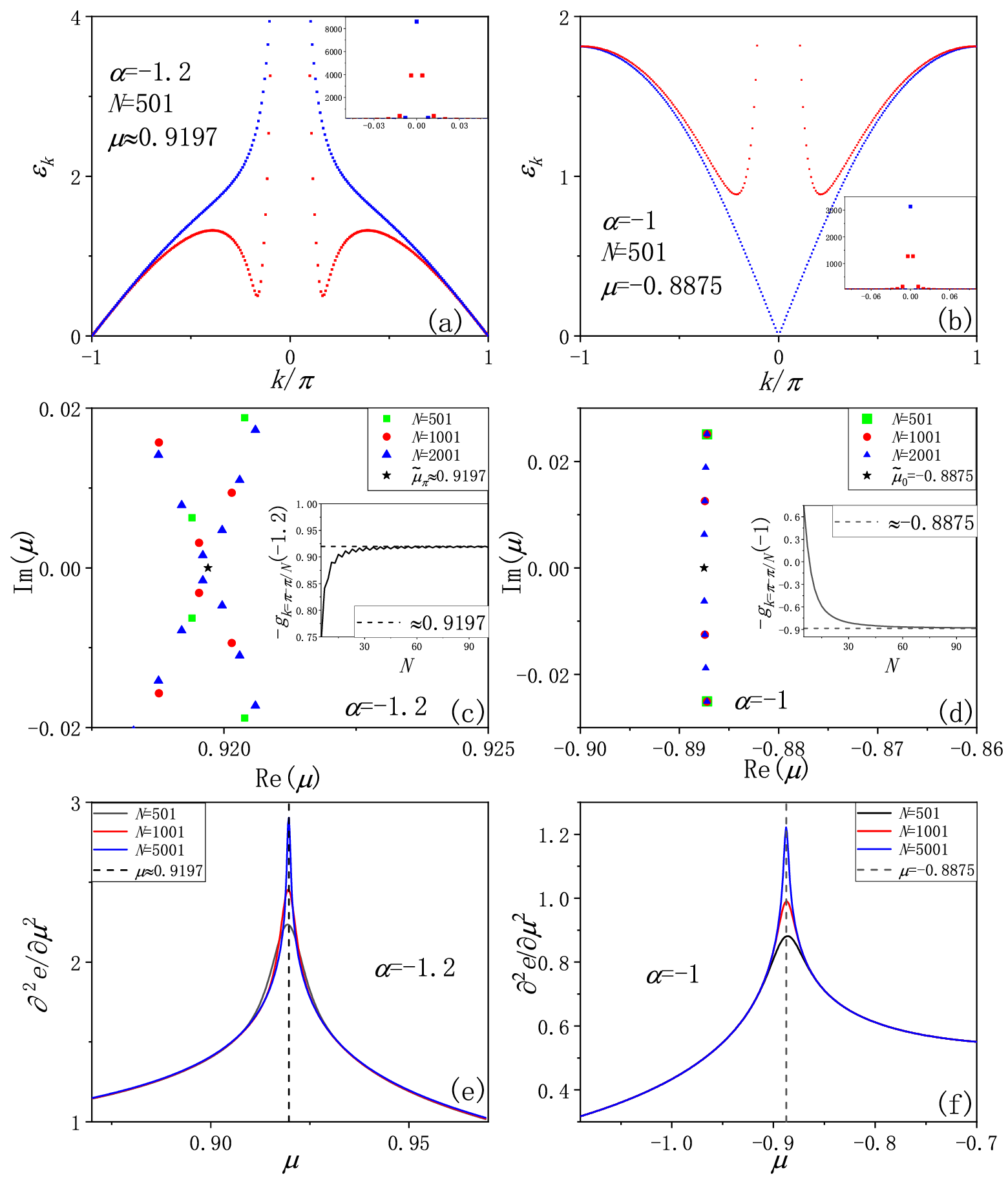}
 \caption{\label{5} (Color online) (a) $\mu\approx0.9197\cdots$ and (b) $\mu=-0.8875$: The QES for $N=501$ sites.
 (c)-(d) The zeros are distributed near the QPT points.
 The insets show the real part of zeros $-g_{k=\pi-\pi/N}(-1.2)$ and $-g_{k=4\pi/N}(-1)$ as functions of $N$, respectively.
 (e)-(f) The second order partial derivative of the ground-state energy density as functions of $\mu$.
 (a), (c), and (e) $\alpha=-1.2$.
 (b), (d), and (f) $\alpha=-1$.
 }
\end{figure}

\emph{The case of $-2<\alpha<0$}.---
In Figs. \ref{1}(d)-(f), we show the zeros in the complex $\mu$ plane for three typical $\alpha$ in this regime.
Quite different from those in the case of $\alpha\geq0$, the zeros constitute two curves.
Meanwhile, the QES is discontinuous functions of the wave vector $k$ and have two branches.
Two typical examples of QES are shown in Figs. \ref{5}(a)-(b).
One branch corresponds to $k=2n\cdot2\pi/N$ (blue points) and the other to $k=(2n-1)\cdot2\pi/N$ (red points).
This is due to the summation
$\sum_{m=1}^{[N/2]}\frac{f}{m^\alpha}\cos(mk)$ in the $\varepsilon_k$ being a discontinue function of $k$ \cite{Sup}.
In the limit $k\rightarrow\pi$ $(N\rightarrow\infty)$, the zeros in two curves
converge to the same $\tilde{\mu}_{\pi}(\alpha)$ on the real axis.
As a typical example, we plot the zeros around $\tilde{\mu}_\pi(-1.2)=0.9197\cdots$ for different $N$ in Fig. \ref{5}(c).
Obviously, the points closest to the real axis of the two curves approach $\tilde{\mu}_{\pi}(-1.2)$ with increasing $N$ (see the inset).
Similarly, the second order partial derivation of the ground-state energy density shows that the QPTs are second order [see Fig. \ref{5}(e)].
This means that there is a QPT point at $\mu=\tilde{\mu}_{\pi}(\alpha)$ in the system of $-2<\alpha<0$ \cite{Sup}.

When $-2<\alpha <0$, the $\psi(\alpha,\infty)$ are divergent.
One cannot obtain the $\mu_{\pi}(\alpha)$ by taking direct summation.
Likewise, we can define the resummation $\psi(\alpha,\infty)$ by analytic continuation of $\eta(\alpha)$ \cite{Sup,Edw,Tit2}.
For example, the $\psi(-1.2,\infty):=\eta(-1.2)=0.1969\cdots$.
Correspondingly, the $\mu_\pi(-1.2):=0.9196\cdots$ which is equal to $\tilde{\mu}_\pi(-1.2)$ obtained numerically.
This is similar to that in the case of $\alpha=0$, the same situation holds for every $-2<\alpha<0$. Consequently, for $-2<\alpha<0$, there is a QPT point at $\mu=\mu_{\pi}(\alpha)$ corresponding to the resummation of divergent series.

On the other hand, the situation of zeros corresponding to $k\rightarrow0$ is different from that of $k\rightarrow\pi$.
From the Figs. \ref{1}(d)-(f), it can be seen that the zeros labeled by red points [corresponding to $k=(2n-1)\cdot2\pi/N$] always go to $+\infty$ while blue points (corresponding to $k=2n\cdot2\pi/N$) tend to $+\infty$ ($-\infty$) for $-1<\alpha<0$ ($\alpha<-1$).
However, for $\alpha=-1$, the blue zeros tend to $\tilde{\mu_0} (-1)=-0.8875\cdots$ [see in Fig. \ref{5}(d) and its inset], which can be obtained by the summation $\sum_{m=1}^{\infty}m\cos(\frac{4m\pi}{N})=-1/8$ \cite{Sup}.
But it does not equal the value $\zeta(-1):=-1/12$ obtained by the analytic continuation.


Moreover, we discuss the topological properties.
Due to the discontinuity of Bogoliubov angle $\theta_{k}$ \cite{Sup} as a function of $k$, the winding number cannot be calculated by direct integration. Instead, we can obtain it by using the integral of the distribution \cite{Bar}.
It is found that the system exists two novel topological phases.
In these phases, the winding numbers are dependent on the parity of $[N/2]$ for finite $N$.
When $N\rightarrow\infty$, the winding number is similar to the resummation of $\pm\sum_{m=1}^{\infty}(-1)^{m-1}$ which can be defined by analytic continuation of $\pm\eta(0):=\pm1/2$ \cite{Sup}.
The results are briefly summarised as follows (see the details in the Supplemental Material \cite{Sup}):
($\rm{\romannumeral1}$) when $-1<\alpha<0$, the phase is topologically trivial for $\mu<\mu_\pi(\alpha)$ and nontrivial with $\omega=-1$ for $\mu>\mu_\pi(\alpha)$.
($\rm{\romannumeral2}$) When $\alpha=-1$, the winding numbers can be defined as $0$, $+1/2$, and $-1/2$ for $\mu<\mu_0(-1)$, $\mu_0(-1)<\mu<\mu_\pi(-1)$, and $\mu>\mu_\pi(-1)$.
($\rm{\romannumeral3}$) When $-2<\alpha<-1$, $\omega:=+1/2$ for $\mu<\mu_\pi(\alpha)$ and $\omega:=-1/2$ for $\mu>\mu_\pi(\alpha)$.
Note that although there are topological phases with half-integer winging numbers, the physical origin is different from that in LR pairing Kitaev model \cite{Pez,Viy,Sup}.

Similarly, we also study the von Neumann entropy in this case.
The properties of entropy are quite different from that in the case of $\alpha\geq0$.
The entropy of a subchain with length $L$ is $S(L)=A(L-L_0)^{\beta}$ no matter in the gapped region or not (see the details in the Supplemental Material \cite{Sup}).
And the exponent $0<\beta<1$ and $\beta\approx1$ for $\alpha>-1$ and $-2<\alpha\le-1$, respectively.

\emph{The case of $\alpha\le-2$}.---
Different from that in the case of $-2<\alpha<0$, where the zeros of the two branches tend to same point in real axis as $k\rightarrow \pi$, the zeros tend to two different points for $\alpha=-2$ as $N\rightarrow \infty$ (see the details in the VIII.A of \cite{Sup}).
Therefore, there are no convergent QPT points in the thermodynamic limit.
When $\alpha<-2$, the zeros in one branch go to $-\infty$, while the other tend toward $+\infty$ \cite{Sup}.
So there are no QPT points in the system when $\alpha\le2$.

It is also found that the winding number equals to 0 for $\alpha\le-2$, so the phase is topologically trivial.
However, the von Neumann entropy of a subchain with length $L$ is $S(L)=A(L-L_0)^{\beta}$.
And $\beta\approx1$ which is similar to that in the case of $-2<\alpha\le-1$. (see the details in the Supplemental Material \cite{Sup})

\emph{Conclusion and discussion}.---In this letter we extend the Kitaev chains for fermions with infinite-range ($\alpha=0$) or divergent LR ($\alpha<0$) hopping.
By examining the Lee-Yang zeros in the complex $\mu$ plane, we found that there are QPT points corresponding to the resummation of divergent series $\psi(\alpha,\infty)$ for $-2<\alpha\le0$.
Simultaneously, there are two isolated QPT points at $\mu=\mu_0(\alpha=0)$ and $\mu_0(\alpha=-1)$.
The $\mu_0(\alpha=0)$ corresponds to the resummation of divergent series $\varphi(0,\infty)$ while the $\mu_0(\alpha=-1)$ doesn't correspond to the resummation of divergent series $\varphi(-1,\infty)$.
Meanwhile, for $\alpha<0$, the QES divides into two branches which is quite different from that in the SR or LR systems with $\alpha\ge0$.
This results in the occurrence of the von Neumann entropy with power law of the subchain length and the topological phases with half-integer winding numbers.
Our main results are summarized in Fig. \ref{ph} and table $\rm{\uppercase\expandafter{\romannumeral1}}$.

Although our studies are focused on the extended Kitaev chains with LR hopping, it can be generalized to the Kitaev chains with long-range pairing \cite{Vod}, the spin-1/2 chains with multispin interactions \cite{Zha}, and other systems with LR interactions \cite{Vod2}.
These studies provide examples for the application of divergent series in condensed matter physics.
Finally, we expect these results can be verified by trapped ion \cite{Dou,Bri,Fri,Kim}, cold atoms \cite{Sch,Nev}, and quantum simulator platform \cite{Per}.

\emph{Acknowledgments}.---The authors thank Chaoqiang Geng, Guoxiong Jin, Leihan Tang, Huicheng Yin, and Gao Zhang for helpful discussions.
We acknowledge the support of the National Natural Science Foundation of China (Grant Nos. 12247106, 12404219, and 11975126).

\begin{table}[htb]
\center
\item Table $\rm{\uppercase\expandafter{\romannumeral1}}$.
The relationship between QPTs and $\varphi(\alpha,\infty)$ or $\psi(\alpha,\infty)$ for different $\alpha$.
\begin{tabular}{|c|c|c|c|c|c|c|}\hline
\multicolumn{1}{|c|}{\multirow{2}{*}{$\alpha$}} & \multicolumn{3}{c|}{$\varphi(\alpha,\infty)$} & \multicolumn{3}{c|}{$\psi(\alpha,\infty)$} \\ \cline{2-7}
& \scriptsize{C} or \scriptsize{D} & \scriptsize{QPT} & $S(L)\sim$ & \scriptsize{C} or \scriptsize{D} & \scriptsize{QPT} & $S(L)\sim$ \\ \hline
$\alpha\ge2$ & C & Y & $a\log L$ & C & Y & $a\log L$ \\ \hline
$1<\alpha<2$ & C & Y & $b\log L$ & C & Y & $a\log L$ \\ \hline
$0<\alpha\le1$ & D & N & \diagbox{}{} & C & Y & $a\log L$ \\ \hline
$\alpha=0$ & D & Y & $a\log L$ & D & Y & $a\log L$ \\ \hline
\makecell[c]{$-2<\alpha<0$\\$(\alpha\ne-1)$} & D & N & \diagbox{}{} & D & Y & $(L-L_0)^\beta$ \\ \hline
$\alpha=-1$ & D & Y2 & $(L-L_0)^\beta$ & D & Y & $(L-L_0)^\beta$ \\ \hline
$\alpha\le-2$ & D & N & \diagbox{}{} & D & N & \diagbox{}{} \\ \hline
\end{tabular}
\item Here the C and D represent convergent and divergent series, respectively.
Y represents that the QPT point converge to the value which corresponds to $\zeta$ or $\eta$ functions.
Y2 represents that the QPT point converge to the value which doesn't correspond to $\zeta$ functions.
N represents that there are no QPT points.
$a=1/6$, $b\in[0,1/6]$, and $\beta\in[0,1]$.
\end{table}

\bibliographystyle{apsrev4-1}
\bibliography{test}

\end{document}


\renewcommand{\baselinestretch}{2}
 \baselineskip 24pt
\title{Supplemental Material: \\ Resummation of divergent series and quantum phase transitions in Kitaev chains with long-range hopping}
\author{Hao Fu$^{1}$}
\noaffiliation
\affiliation{Department of Mathematics and Physics, Bengbu University, Bengbu, 233030, People's Republic of China}
\author{Peiqing Tong$^{2,}$}
\email{pqtong@njnu.edu.cn}
\noaffiliation
\affiliation{Ministy of Education Key Laboratory of NSLSCS and School of Physics and Technology, Nanjing Normal University, Nanjing, 210023, People's Republic of China}
\date{\today}

\maketitle
\tableofcontents
\section{Models and Diagonalization}
We start from the Kitaev chain with long-range (LR) hopping whose Hamiltonian is
\begin{equation}\label{ham}
\begin{split}
  H=-\sum_{n=1}^{N}[tc_n^\dagger c_{n+1}+\Delta c_n^\dagger c_{n+1}^\dagger+
  \mu(c_n^\dagger c_{n}-\frac{1}{2})]
  -\sum_{n=1}^{N}\sum_{m=2}^{[N/2]}\frac{f}{m^\alpha}c_n^\dagger c_{n+m}+{\rm H.C.},
\end{split}
\end{equation}
where $c_n^\dagger$ ($c_n$) is a fermionic creation (annihilation) operator at site $n$, $t$ the strength of short-range (SR) hopping, $\Delta$ the strength of SR pairing, $f$ the strength of LR hopping, $\mu$ the chemical potential, $N$ the length of lattice, and $[N/2]$ the rounding of $N/2$.
We set $t=\Delta=1$ without loss of generality.
By the Fourier transformation
\begin{equation}\nonumber
\begin{split}
  c_n=\frac{1}{N}\sum_{k}e^{-ikn}c_k,
\end{split}
\end{equation}
Hamiltonian (\ref{ham}) can be written as
\begin{equation}\label{hamc}
\begin{split}
  H=\sum_{k}H_k=\sum_{k}\Gamma^{\dagger}M_k\Gamma,
\end{split}
\end{equation}
where
\begin{equation}\label{mk}
\begin{split}
  M_k=\left(\begin{array}{cc}-\mu-g_k(\alpha)&i\sin k\\
  -i\sin k&\mu+g_k(\alpha)
      \end{array}\right)
\end{split}
\end{equation}
with
\begin{equation}\label{g1}
\begin{split}
  g_k(\alpha)=\cos k+\sum_{m=2}^{[N/2]}\frac{f}{m^\alpha}\cos(mk),
\end{split}
\end{equation}
and $\Gamma^{\dagger}=(c_{k}^{\dagger},c_{-k})$.
Under periodic boundary condition, $k=\frac{2\pi}{N}h$ and $h$ is an integer between $-[N/2]$ and $[N/2]$ .

Using the Bogoliubov transformation $c_k=u_k\gamma_k+v_{-k}^{\star}\gamma_{-k}^{\dagger}$, the Hamiltonian (\ref{hamc}) can be expressed as
\begin{equation}\nonumber
\begin{split}
  H=\sum_k\varepsilon_k(2\gamma_k^\dagger\gamma_k-1),
\end{split}
\end{equation}
where
\begin{equation}\label{ek}
\begin{split}
  \varepsilon_k=\sqrt{[g_k(\alpha)+\mu]^2+\sin^2 k},
\end{split}
\end{equation}
$u_k=i\sin\frac{\theta_{k}}{2}$ and $v_k=\cos\frac{\theta_{k}}{2}$ with the Bogoliubov angle $\theta_k=\arctan\frac{\sin k}{\mu+g_k(\alpha)}$.

Another form of LR hopping is $\sum_{n=1}^{N}\sum_{m=2}^{N-1}\frac{f}{d_m^\alpha}(c_n^\dagger c_{n+m}+h.c.)$ \cite{Vod2,Pat,Ale,Jag,Rad}, where $d_m=m$ $(d_m=N-m)$ if $m\le N/2$ $(m>N/2)$. The Hamiltonian becomes
\begin{equation}
\begin{split}\label{eq2}
  \tilde{H}=-\sum_{n=1}^{N}[tc_n^\dagger c_{n+1}+\Delta c_n^\dagger c_{n+1}^\dagger+
  \mu(c_n^\dagger c_{n}-\frac{1}{2})]
  -\sum_{n=1}^{N}\sum_{m=2}^{N-1}\frac{f}{d_m^\alpha}c_n^\dagger c_{n+m}+{\rm H.C.}.
\end{split}
\end{equation}
Similarly, the quasiparticle energy spectrum (QES) of Hamiltonian (\ref{eq2}) is $\varepsilon_k=\sqrt{[\tilde {g}_k(\alpha)+\mu]^2+\sin^2 k}$ with
\begin{equation}\nonumber
\begin{split}
 \tilde {g}_k(\alpha)=\cos k+\sum_{m=2}^{N-1}\frac{f}{d_m^{\alpha}}\cos(km).
\end{split}
\end{equation}
In this letter we only discuss the case of odd $N$, so $\tilde {g}_k(\alpha)$ yields
\begin{equation}\nonumber
\begin{split}
 \tilde{g}_k(\alpha)&=\cos k+\sum_{m=2}^{[N/2]}\frac{f}{m^{\alpha}}\cos(km)+\sum_{m=[N/2]+1}^{N-1}\frac{f}{(N-m)^{\alpha}}\cos(km)\\
 &=\cos k+\sum_{m=2}^{[N/2]}\frac{f}{m^{\alpha}}\cos(km)+\sum_{l=1}^{[N/2]}\frac{f}{l^{\alpha}}\cos[k(N-l)].
\end{split}
\end{equation}
For $k=\frac{2\pi}{N}h$, $\cos[k(N-l)]=\cos(kl)$. Then
\begin{equation}\nonumber
\begin{split}
 \tilde{g}_k(\alpha)&=(1+f)\cos k+2\sum_{m=2}^{[N/2]}\frac{f}{m^{\alpha}}\cos(km).
\end{split}
\end{equation}
which is similar to Eq. (\ref{g1}).
Therefore, we choose to use the LR hopping term $\sum_{n=1}^{N}\sum_{m=2}^{[N/2]}\frac{f}{m^\alpha}(c_n^\dagger c_{n+m}+h.c.)$ in the letter.

\section{Resummation of Divergent Series}
\begin{figure}
 \centering
\includegraphics[width=.9\linewidth]{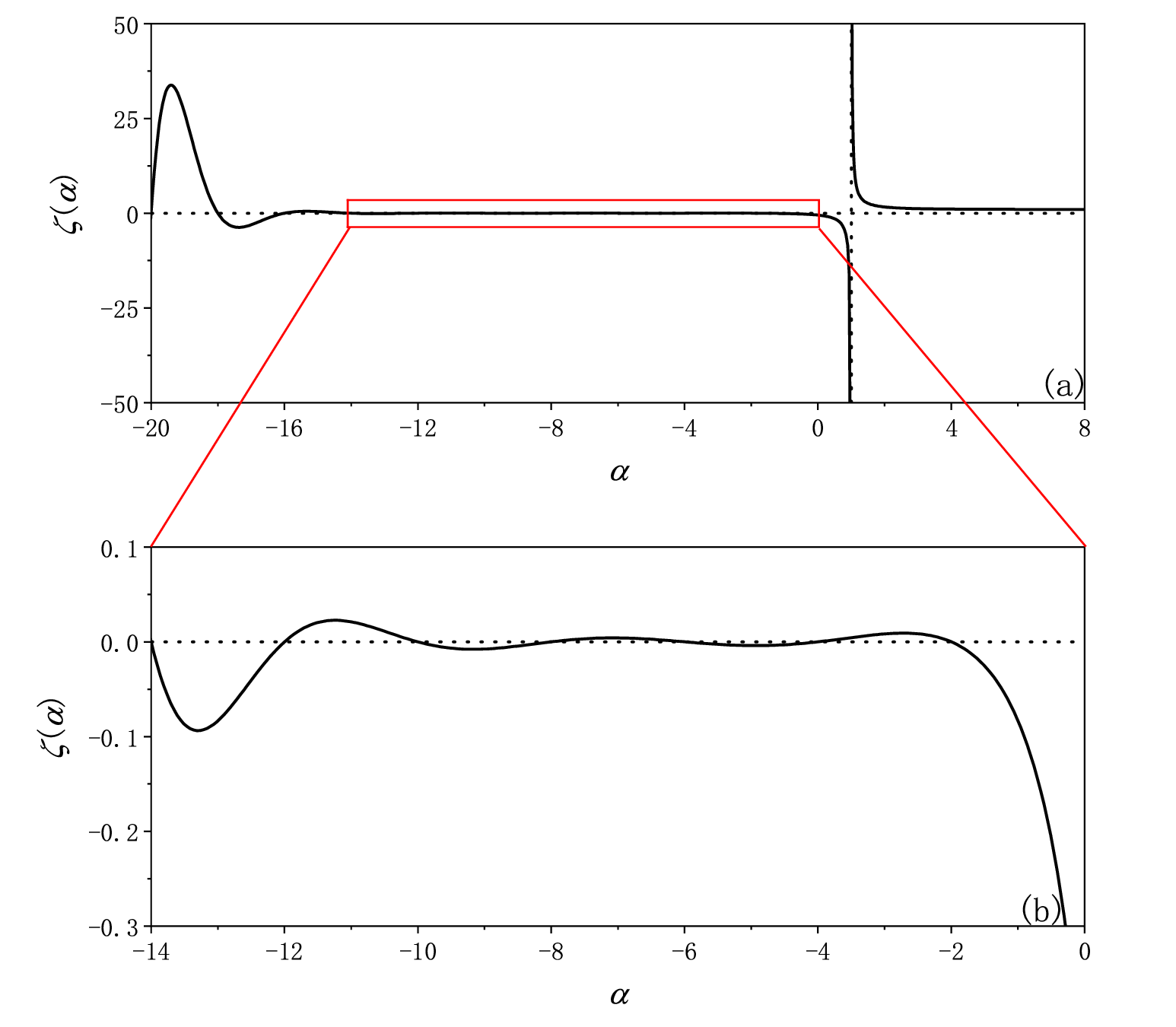}
 \caption{\label{ze} (Color online) The $\zeta(\alpha)$ as a function of real $\alpha$ with different scales.}
\end{figure}

The Riemann zeta function $\zeta(\alpha)$
is very important in number theory \cite{Edw,Tit2} which reads
\begin{equation}\nonumber
  \zeta(\alpha)=\sum_{m=1}^{\infty}\frac{1}{m^{\alpha}},\quad{\rm Re}(\alpha)>1.
\end{equation}
However, it is proven \cite{Edw,Tit2} that the $\zeta(\alpha)=\sum m^{-\alpha}$ can be continued analytically beyond the half-plane Re$(\alpha)> 1$.
By using the analytic continuation, we can obtain the $\zeta(\alpha)$ for $\alpha<1$.
In Fig. \ref{ze}, we show the $\zeta(\alpha)$ as a function of real $\alpha$ with different scales \cite{Sup3}.
Therefore, we can define \cite{Edw,Tit2} the resummation $\sum_{m=1}^{\infty} m^{-\alpha}$ for $\alpha<1$ by analytic continuation as
\begin{equation}\nonumber
  \sum_{m=1}^{\infty} m^{-\alpha}:= \zeta(\alpha),
\end{equation}
where the := sign hence can be read as ``the value from analytic continuation equals to".
Therefore, one can obtain some counterintuitive results, such as
\begin{equation}\nonumber
\begin{split}
  &\sum_{m=1}^{\infty}m^0=1+1+1+1+1+\cdots:=\zeta(0)=-\frac{1}{2},\\
  &\sum_{m=1}^{\infty}m=1+2+3+4+5+\cdots:=\zeta(-1)=-\frac{1}{12}.
\end{split}
\end{equation}

\begin{figure}
 \centering
\includegraphics[width=.9\linewidth]{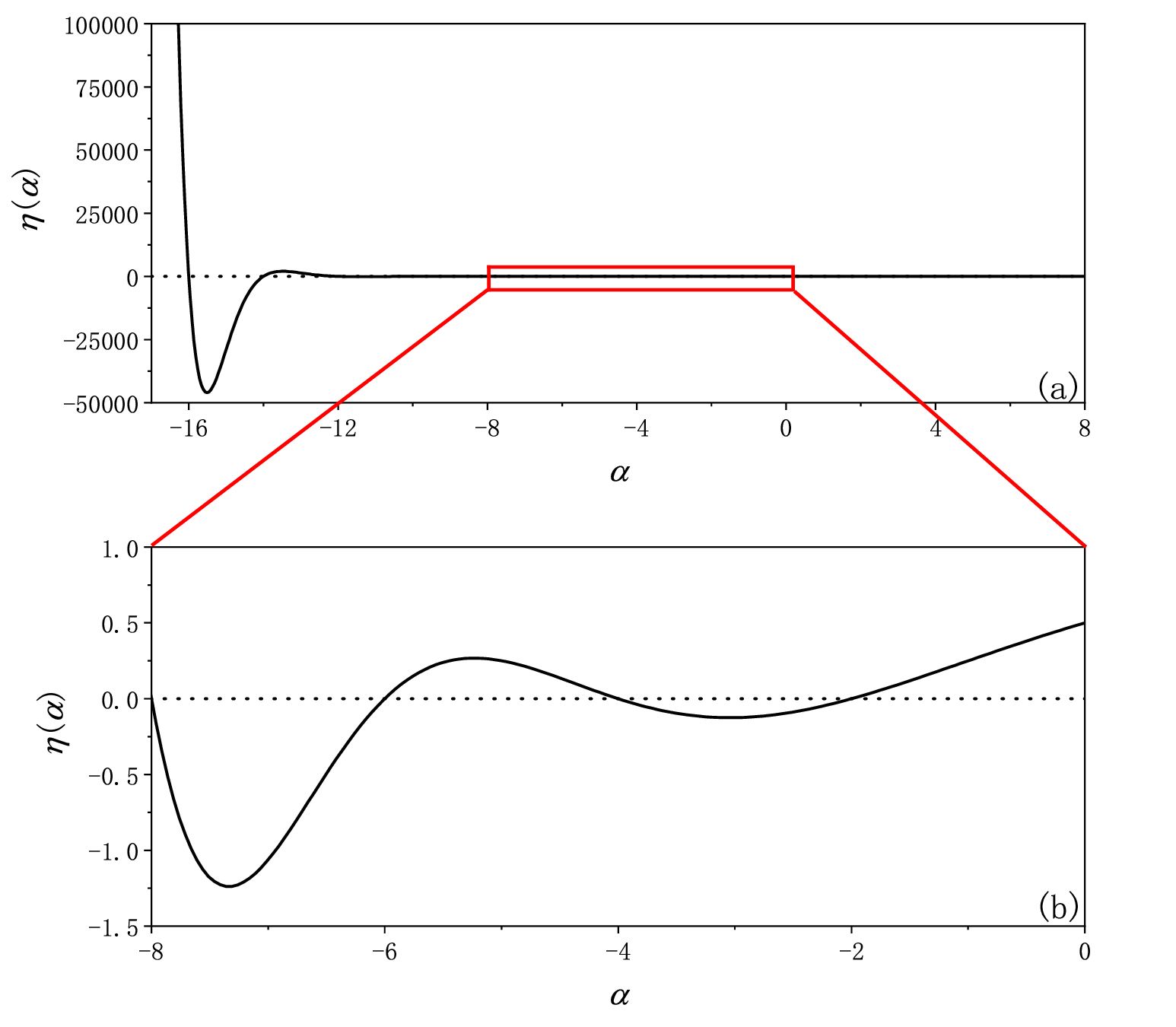}
 \caption{\label{et} (Color online) The $\eta(\alpha)$ as a function of real $\alpha$ with different scales.}
\end{figure}

The Dirichlet eta function, also known as the alternating zeta function, is defined as \cite{Mil}
\begin{equation}\nonumber
  \eta(\alpha)=\sum_{m=1}^{\infty}\frac{(-1)^{m-1}}{m^{\alpha}},\quad{\rm Re}(\alpha)>0.
\end{equation}
It is related to the $\zeta(\alpha)$ by following equation \cite{Mil}
\begin{equation}\nonumber
\eta(\alpha)=(1-\frac{1}{2^{\alpha-1}})\zeta(\alpha).
\end{equation}
Then we can obtain the $\eta(\alpha)$ for $\alpha\le0$ and the numerical results are shown in Fig. \ref{et} \cite{Sup3}.

Similarly, for $\alpha\leq 0$, we can define the resummation of divergent series as
\begin{equation}\nonumber
\sum_{m=1}^{\infty}\frac{(-1)^{m-1}}{m^{\alpha}}:=\eta(\alpha).
\end{equation}
Therefore,
\begin{equation}\nonumber
\begin{split}
  &\sum_{m=1}^{\infty}(-1)^{m-1}m^{0}=1-1+1-1+1-\cdots:=\eta(0)=\frac{1}{2},\\
  &\sum_{m=1}^{\infty}(-1)^{m-1}m=1-2+3-4+5-\cdots:=\eta(-1)=\frac{1}{4},
\end{split}
\end{equation}
{\sl etc.}

\section{Quasiparticle Energy Spectrum}
\begin{figure}
 \centering
\includegraphics[width=.9\linewidth]{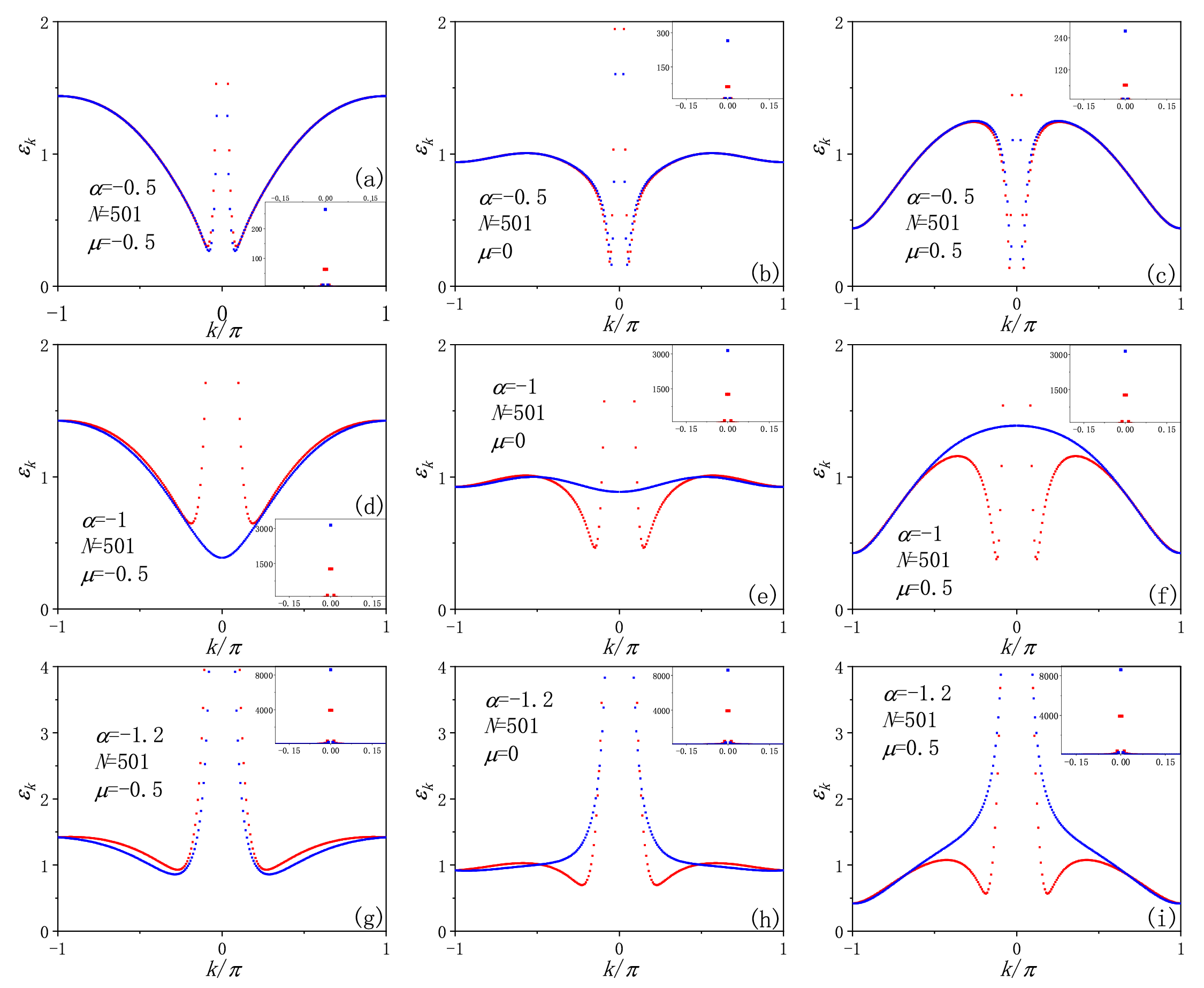}
 \caption{\label{p1} (Color online) The QES $\varepsilon_k$ as functions of $k$ for various $\alpha$ and $\mu$.
  The red and blue points correspond to $k=(2n-1)\cdot2\pi/N$ and $k=2n\cdot2\pi/N$, respectively.
  The high energy spectrum are shown in the insets.}
\end{figure}

\begin{figure}
 \centering
\includegraphics[width=.9\linewidth]{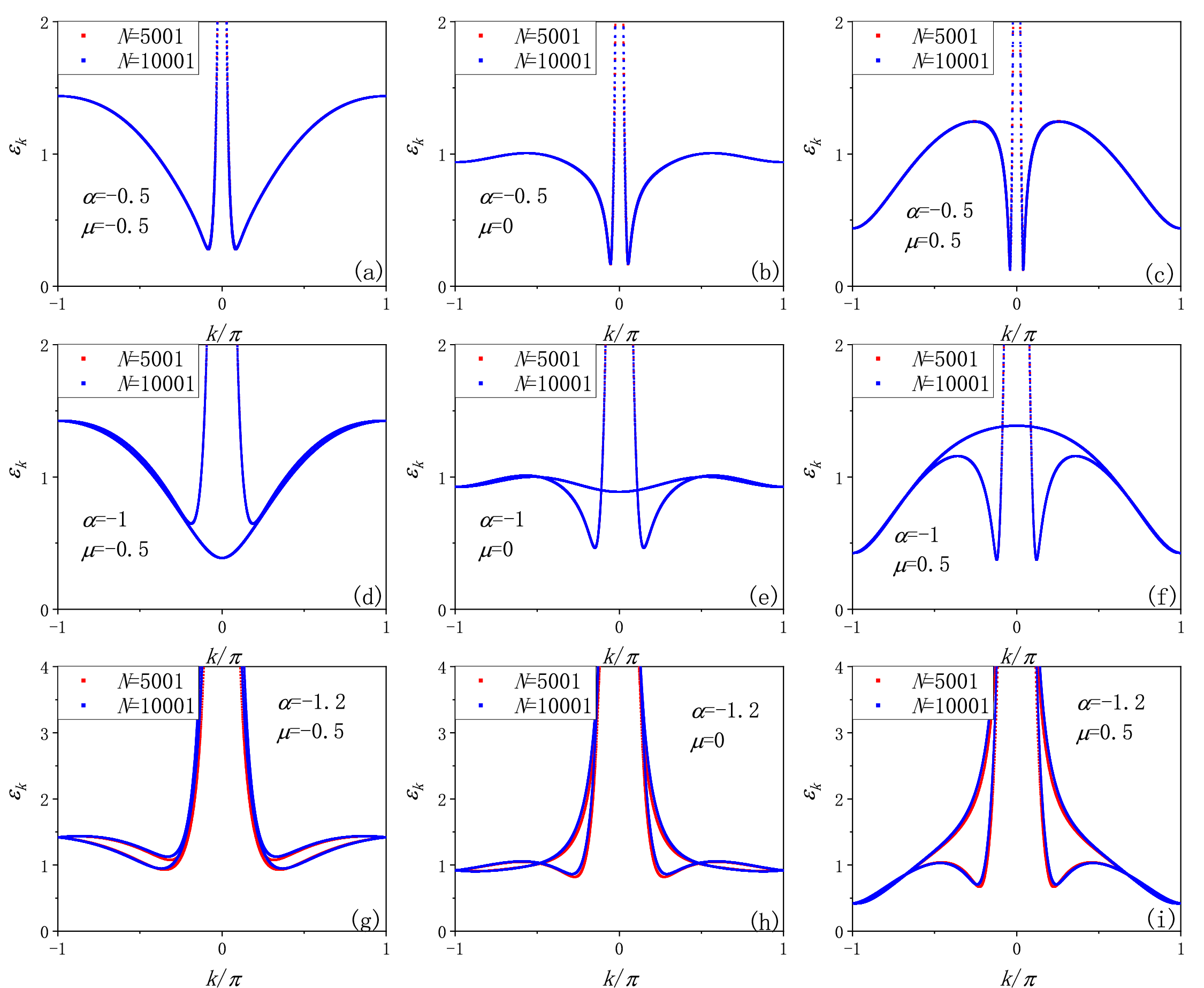}
 \caption{\label{a12} (Color online) The QES $\varepsilon_k$ as functions of $k$ for various $\alpha$ and $\mu$.
 The red and blue points correspond to $N=5001$ and $10001$, respectively.}
\end{figure}

In this section, we mainly study the quasiparticle energy spectrum (QES) for $\alpha<0$.
In Fig. \ref{p1} we give some typical examples for various $\alpha$ and $\mu$ at $N=501$.
The blue and red points in these figures correspond to $k=2n\cdot2\pi/N$ and $k=(2n-1)\cdot2\pi/N$ with $n$ the integer, respectively.
From these figures, we can find that $\varepsilon_k$ is not a continuous function of $k$, {\sl i.e.}, discontinuous for adjacent $k$.
More accurately, the QES splits into two branches (except one point corresponding to $k=0$).
One corresponds to $k=2n\cdot2\pi/N$ and the other corresponds to $k=(2n-1)\cdot2\pi/N$.
In addition, the two branches coincide at $k\rightarrow\pi$.
To discuss the finite size effect on the QES, the QES for different $N$ are shown in Fig. \ref{a12}.
From them we can find that the spectra are almost the same for large $N$.
So the QES for $\alpha<0$ is quite different from that in the case of $\alpha\ge0$.
The corresponding Lee-Yang zeros [as depicted in Fig. 1(d)-(f) in the letter for $\alpha<0$] are also composed of two branches.

\begin{figure}
 \centering
\includegraphics[width=.9\linewidth]{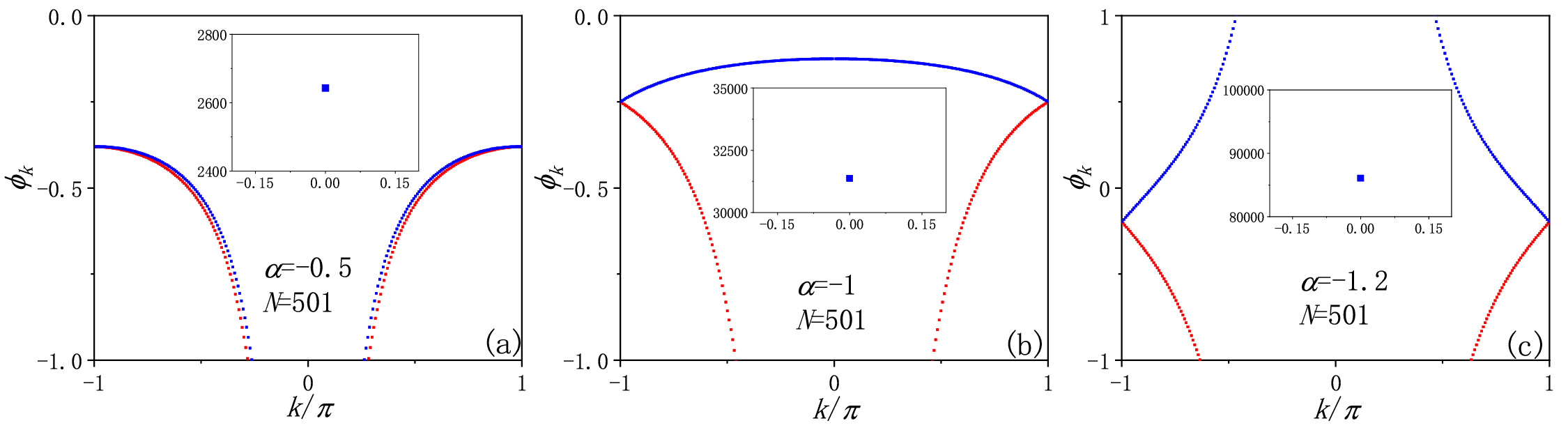}
 \caption{\label{p2} (Color online) The functions $\phi_k$ with respect to $k$ for various $\alpha$.
  The blue and red points correspond to $k=2n\cdot2\pi/N$ and $k=(2n-1)\cdot2\pi/N$, respectively. The insets are the $\phi_k$ of $k=0$.}
\end{figure}

In order to understand the properties of QES for $\alpha<0$, we study the function
\begin{equation}
\begin{split}\label{phi}
  \phi_{k}(\alpha)=\sum_{m=1}^{[N/2]}\frac{1}{m^\alpha}\cos(mk).
\end{split}
\end{equation}
Fig. \ref{p2}(a)-(c) show $\phi_k$ as functions of $k$ for $\alpha=-0.5$, $-1$, and $-1.2$, respectively.
Similarly, the blue and red points correspond to $k=2n\cdot2\pi/N$ and $k=(2n-1)\cdot2\pi/N$, respectively.
From these figures, we can find that $\phi_k$ is discontinuous.

In additions, we can obtain $\phi_k(\alpha)$ for $\alpha=-1$ analytically.
For odd $N$, the Eq. (\ref{phi}) can be expressed  as
\begin{equation}\nonumber
\begin{split}
  \phi_{k}(-1)&=\sum_{m=1}^{[N/2]}m\cos(mk)=\frac{{\rm d}}{{\rm d}k}\sum_{m=1}^{[N/2]}\sin(mk)=\frac{{\rm d}}{{\rm d}k}\frac{\cos(k/2)-\cos(Nk/2)}{2\sin(k/2)}\\
  &=\frac{\sin(k/2)[N\sin(Nk/2)-\sin(k/2)]-\cos(k/2)[\cos(k/2)-\cos(Nk/2)]}{4\sin^2(k/2)}\\
  &=\frac{N\sin(k/2)\sin(Nk/2)+\cos(k/2)\cos(Nk/2)-1}{4\sin^2(k/2)}.
\end{split}
\end{equation}
When $k=\frac{2\pi}{N}h$,
\begin{equation}
\begin{split}\label{phi3}
  \phi_{k}(-1)\equiv\phi_{h}(-1)=\frac{\cos(h\pi/N)\cos(h\pi)-1}{4\sin^2(h\pi/N)}.
\end{split}
\end{equation}
Hence,
\begin{equation}\nonumber
\phi_{h+1}(-1)-\phi_{h}(-1)=\left\{\begin{array}{c} u_h+v_h\qquad {\rm odd} \quad h,\\
 -u_h+v_h \qquad {\rm even} \quad h,\end{array}\right.
\end{equation}
here
\begin{equation}
\begin{split}\nonumber
u_h=&\frac{\cos[(h+1)\pi/N]}{4\sin^2[(h+1)\pi/N]}+\frac{\cos(h\pi/N)}{4\sin^2(h\pi/N)}\\
v_h=&-\frac{1}{4\sin^2[(h+1)\pi/N])}+\frac{1}{4\sin^2(h\pi/N)}.
\end{split}
\end{equation}
In the limit $N\rightarrow+\infty$, the $v_h$ goes to zero except $h$ closing to $0$ and $u_h$ goes to a finite value except $h$ closing to $0$ and $N/2$. Therefore, $\phi_{k}(-1)$ is discontinuous.

In short, the discontinuity of the QES is due to $\phi_k(\alpha)$.

\section{The Case of $\alpha>0$}
\begin{figure}[t]
 \centering
\includegraphics[width=.9\linewidth]{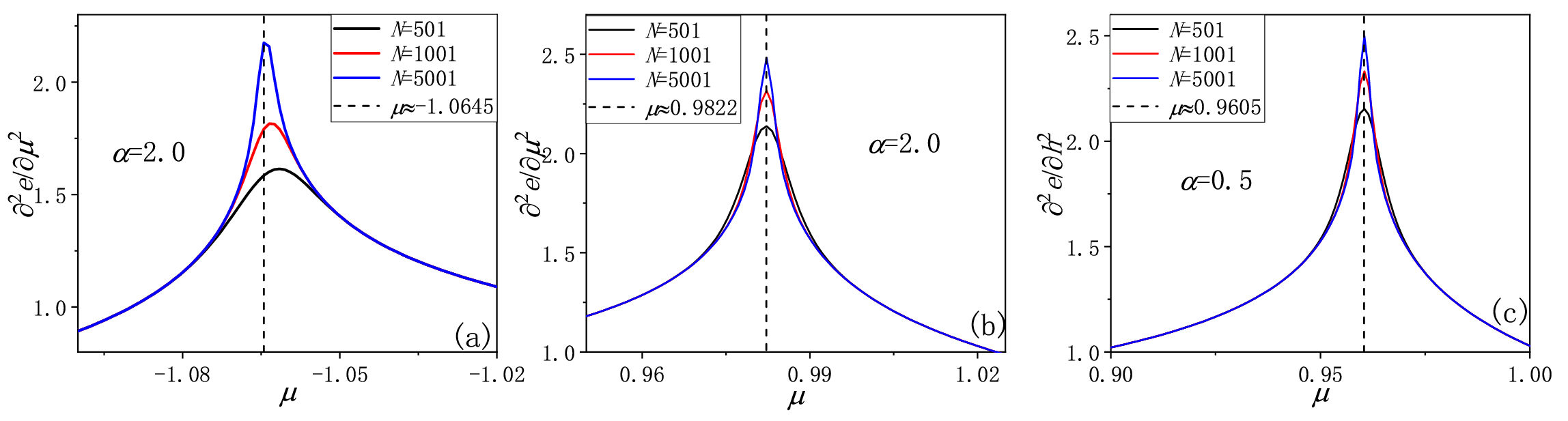}
 \caption{\label{a2e} (Color online) The second order partial derivative of ground-state energy density with respect to $\mu$.
 (a)-(b) $\alpha=2$. (c) $\alpha=0.5$.
 }
\end{figure}
\subsection{Quantum phase transitions}
In this subsection we discuss the quantum phase transition (QPT) points in two cases of $\alpha>1$ and $0<\alpha\le1$, respectively.
When $\alpha>1$, both the zeta and eta functions are convergent.
So the QPT points can be determined by using Eqs. (5a)-(5b).
For examples, the QPT points at $\mu=\mu_0(\alpha)\approx-1.0645$ and $\mu_\pi(\alpha)\approx0.9822$ for $\alpha=2$.
The second-order partial derivative of the ground-state energy density around $\mu=\mu_0(2)$ and $\mu_\pi(2)$ are shown in Fig. \ref{a2e}(a)-(b), respectively.
From the figures, it is clear that the $\frac{\partial^2e}{\partial \mu^2}$ at the QPT point tends to infinity as $N\rightarrow+\infty$.
Therefore, the QPTs are second order.

However, for $0<\alpha\le1$, the zeta function becomes divergent and eta function is still convergent.
Hence, there is only one QPT point which corresponds to $\mu_\pi$.
As an example, $\mu_\pi(\alpha)\approx0.9605$ at $\alpha=0.5$.
Similarly, the $\frac{\partial^2e}{\partial \mu^2}$ [see Fig. \ref{a2e}(c)] tends to infinity at $\mu=\mu_\pi(\alpha)$ for $N\rightarrow+\infty$.

\subsection{Winding number}
In this subsection, we study the topological properties by calculating the winding number, a topological invariant in our model.
For the momentum $k$ going across the whole Brillouin zone from $-\pi$ to $+\pi$, the winding number $\omega$ can be defined as the angle swept by the winding vector $\bm{r}_k=(0,-\sin k,-\mu-g_k(\alpha))$, {\sl i.e.}
\begin{equation}\label{w}
\begin{split}
  \omega\equiv\frac{1}{2\pi}\int_{-\pi}^{\pi} {\rm d}\theta_k,
\end{split}
\end{equation}
where $\theta_k=\arctan\frac{\sin k}{\mu+g_k(\alpha)}$ is the Bogoliubov angle.

\begin{figure}[t]
 \centering
\includegraphics[width=.9\linewidth]{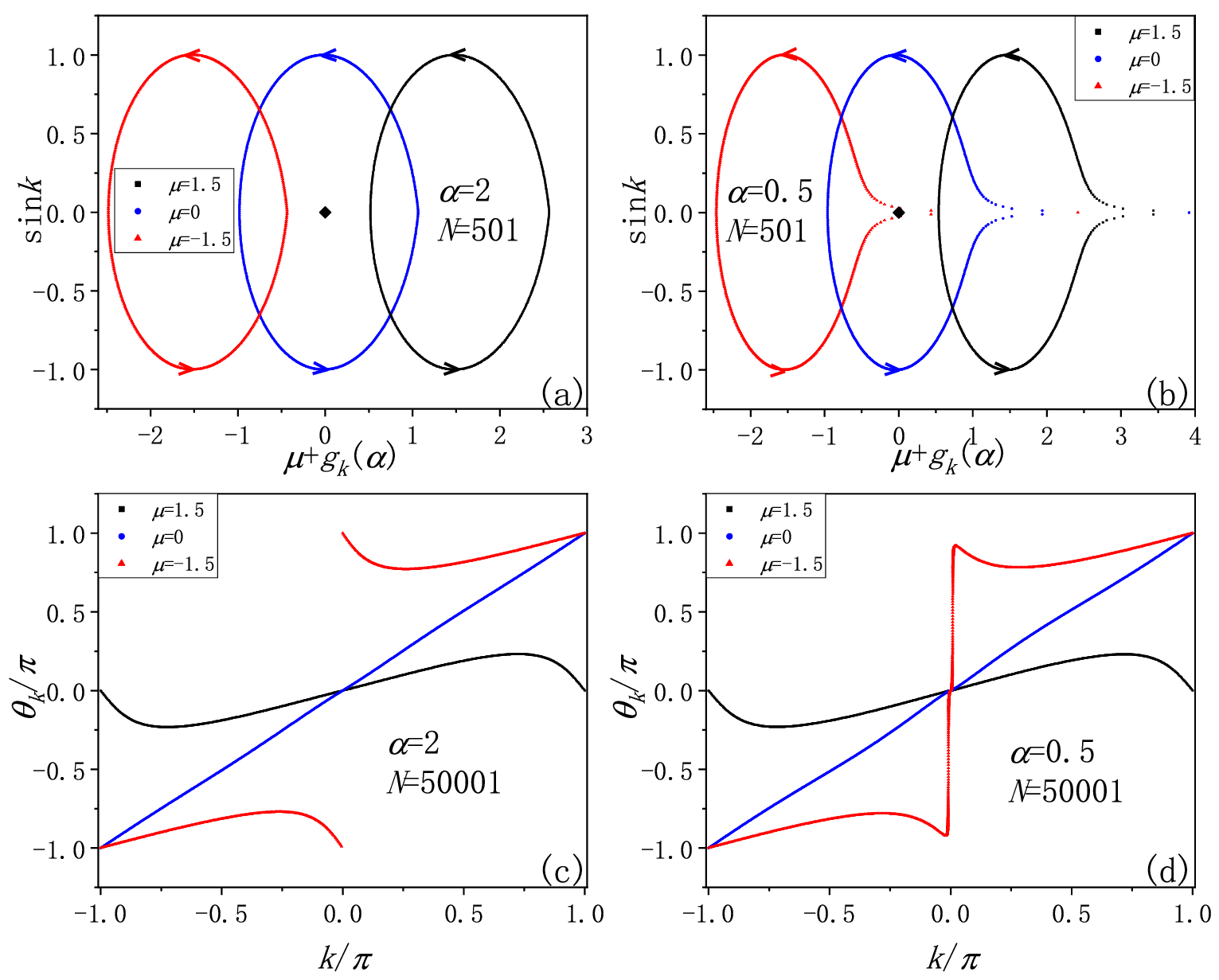}
 \caption{\label{a2w} (Color online) (a)-(b) The winding vectors in the $y-z$ plane corresponding to $\alpha=2$ and $0.5$, respectively.
 (c)-(d) The Bogoliubov angle $\theta_k$ as functions of $k$ corresponding to $\alpha=2$ and $0.5$, respectively.
 The black square point, blue circular point, and red triangular point correspond to $\mu=1.5$, $0$, and $-1.5$, respectively.}
\end{figure}

When $\alpha>1$ ,there are two QPT points at $\mu=\mu_0(\alpha)$ and $\mu_\pi(\alpha)$. 
The winding vector and corresponding Bogoliubov angle $\theta_k$ for $\alpha=2$ are shown in Fig. \ref{a2w}(a) and (c), respectively.
From Fig. \ref{a2w}(a), we can find that the winding vector corresponding to $\mu>\mu_\pi(\alpha)$ (black points) does not wind around the origin in the $y-z$ plane.
From Fig. \ref{a2w}(c), when $k$ is varied from $-\pi$ to $+\pi$, the corresponding Bogoliubov angle (black points) continuously changes from $0$ to $0$.
So the winding number $\omega=0$.
Similarly, the winding number $\omega=0$ also for $\mu<\mu_0(\alpha)$ [see the red points in Fig. \ref{a2w}(a) and (c)].
However, the winding vector corresponding to $\mu_0(\alpha)<\mu<\mu_\pi(\alpha)$ in Fig. \ref{a2w}(a) (blue points) winds around the origin counterclockwise in the $y-z$ plane.
From Fig. \ref{a2w}(c), the corresponding Bogoliubov angle continuously changes from $-\pi$ to $+\pi$.
So the winding number $\omega=+1$.
Therefore, the phase is topologically trivial for $\mu<\mu_0(\alpha)$ or $\mu>\mu_\pi(\alpha)$ while topologically nontrivial for $\mu_0(\alpha)<\mu<\mu_\pi(\alpha)$.

When $0<\alpha\le1$, there is only one QPT point at $\mu=\mu_\pi(\alpha)$.
The winding vector and Bogoliubov angle $\theta_k$ corresponding to $\mu<\mu_\pi(\alpha)$ and $\mu>\mu_\pi(\alpha)$ are shown in Fig. \ref{a2w}(b) and (d), respectively.
In this two figures, we can obtain $\omega=0$ for $\mu>\mu_\pi(\alpha)$ (black points) and $\omega=+1$ for $\mu<\mu_\pi(\alpha)$ (blue and red points).
Different from that in the case of $\alpha>1$, the winding vector tends to positive infinity for $k\rightarrow0$ so the loop of the winding vector always encloses the origin for $\mu<\mu_\pi(\alpha)$.
Therefore, the phase is topologically trivial for $\mu>\mu_\pi(\alpha)$ while topologically nontrivial for $\mu<\mu_\pi(\alpha)$.

\subsection{von Neumann entropy}
The von Neumann entropy measures the entanglement between a block of $L$ contiguous sites (the subchain within a chain of $N$ sites) and the rest of the chain, which is defined as \cite{Ami,Eis3}
\begin{equation}\nonumber
\begin{split}
  S_L=-{\rm Tr}(\rho_L\log_2\rho_L).
\end{split}
\end{equation}
Here, $\rho_L$ is the reduced density matrix obtained from the ground-state wave function in which all the sites but the subchain have been traced out, {\sl i.e.}
\begin{equation}\nonumber
\begin{split}
  \rho_L={\rm Tr}_{\bar{L}}[|\psi_0><\psi_0|],
\end{split}
\end{equation}
where $|\psi_0>$ is the ground-state and $\bar{L}$ represents the rest of the chain.
In the conformally invariant models, the subchain entropy satisfies the equation $S(L)=(c_{{\rm eff}}/3)\log L+b$ with $b$ being a nonuniversal constant and $c_{\rm eff}$ the effective central charge \cite{Ami,Eis3}.
In the SR Kitaev chain $c_{\rm eff}=0.5$ on the critical points and $c_{\rm eff}=0$ in the gapped region (off the critical points).

\begin{figure}[t]
 \centering
\includegraphics[width=.9\linewidth]{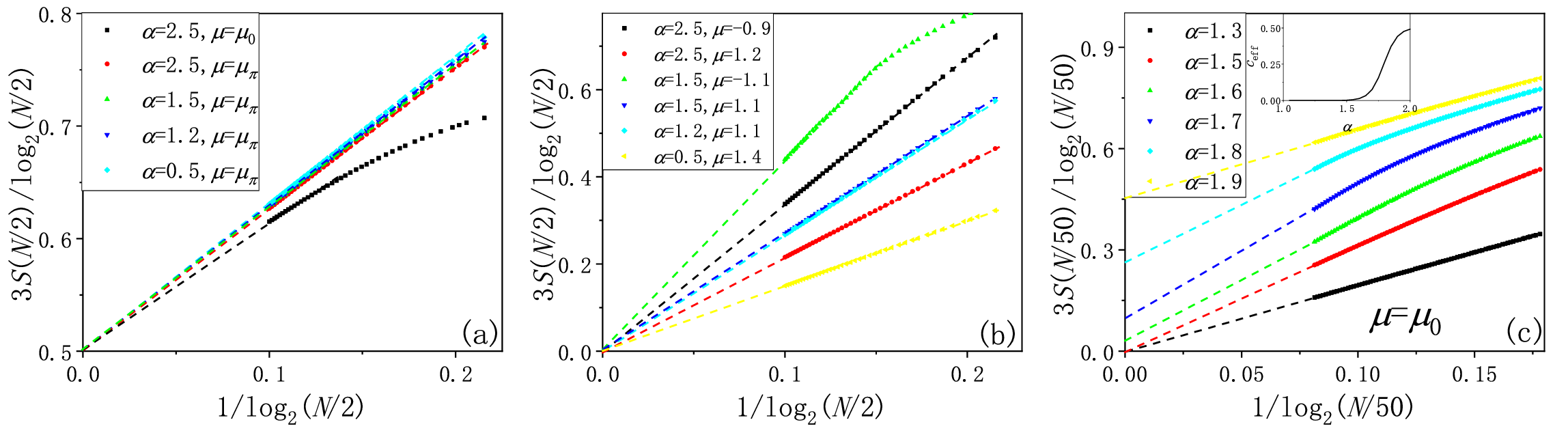}
 \caption{\label{a2s} (Color online) (a) The $3S(N/2)/\log_2(N/2)$ as functions of $1/\log [N/2]$ for various $\alpha$ at QPT points.
 (b) The $3S(N/2)/\log_2(N/2)$ as functions of $1/\log [N/2]$ off the critical points.
 (c) The $3S(N/50)/\log_2(N/50)$ as functions of $1/\log [N/50]$ at the QPT points $\mu_0(\alpha)$ for $1<\alpha<2$.
 Inset: Plot of $c_{{\rm eff}}$ vs $\alpha$ for $\mu=\mu_0(\alpha)$.}
\end{figure}

In this subsection, we study the von Neumann entropy for LR systems for $\alpha>0$.
In order to eliminate the finite size effect, we use the method in Ref. \cite{Vod} to obtain $c_{\rm eff}$ by plotting $3S(N/2)/\log_2(N/2)$ as functions of $1/\log_2(N/2)$.
Fig. \ref{a2s}(a) shows the typical results on the critical line $\mu=\mu_0$ for $\alpha\ge2$ and the critical line $\mu=\mu_\pi$ for $\alpha>0$.
From the figure it is found that the effective central charge $c_{{\rm eff}}=0.5$ on the two critical lines.
In the gapped region, some examples for various $\alpha$ and $\mu$ are shown in Fig. \ref{a2s}(b).
From the figure we find that the effective central charge $c_{{\rm eff}}=0$ in the gapped region for $\alpha>0$.
When $1<\alpha<2$, the convergence of $\mu_0(\alpha)$ is very slow, so the length of lattice $N$ shall be large.
Therefore, we calculate the $S(N/50)/\log_2(N/50)$ with respect to $1/\log_2(N/50)$.
The results are shown in Fig. \ref{a2s}(c). It is found that the effective central charge $0\le c_{{\rm eff}}<0.5$ for $1<\alpha<2$ [see the inset in Fig. \ref{a2s}(c)].

Summarily, the effective central charge yields
\begin{equation}\nonumber
 c_{{\rm eff}}\left\{\begin{array}{c} =0.5,\qquad \mu=\mu_\pi(\alpha),\ \alpha>0,\qquad\\
 [1.5mm] =0.5,\qquad \mu=\mu_0(\alpha),\ \alpha\ge2,\qquad\\
 [1.5mm] \in[0,0.5),\ \mu=\mu_0(\alpha),\ 1<\alpha<2.  \end{array}\right.
\end{equation}
It is worth noting that our results are quite different from those of the Kitaev chains with LR pairing \cite{Vod}.
In that systems, the $c_{\rm eff}>0.5$ on the critical line for $\alpha<1.5$ and $c_{\rm eff}>0$ in gapped region for $\alpha<1$.

To understand the abnormal effective central charge $c_{{\rm eff}}\in[0,0.5)$ at $\mu=\mu_0(\alpha)$ for $1<\alpha<2$, we calculate the derivative of the dispersion relation $\varepsilon'_{0}={\rm d}\varepsilon_{k}/{\rm d}k|_{k\rightarrow0}$.
For $N\rightarrow+\infty$, Eq. (\ref{phi}) can be written as
\begin{equation}\nonumber
\begin{split}
  \phi_{k}(\alpha)=\frac{1}{2}[{\rm Li}_{\alpha}(e^{ik})+{\rm Li}_{\alpha}(e^{-ik})],
\end{split}
\end{equation}
where ${\rm Li}_{\alpha}(z)=\sum_{m=1}^{\infty}\frac{z^m}{m^\alpha}$ is the polylogarithm \cite{Cla}.
Using the following expansion for the polylogarithm:
\begin{equation}\nonumber
\begin{split}
  {\rm Li}_{\alpha}(z)=\Gamma(1-\alpha)(\ln\frac{1}{z})^{\alpha-1}+\sum_{n=0}^{\infty}\zeta(\alpha-n)\frac{\ln^nz}{n!},
\end{split}
\end{equation}
we can get
\begin{equation}\nonumber
\begin{split}
  \phi_{k}(\alpha)=-\Gamma(1-\alpha)\sin(\frac{\alpha\pi}{2})k^{\alpha-1}+\zeta(\alpha)-\frac{1}{2}\zeta(\alpha-2)k^{2}+\mathcal{O}(k^4).
\end{split}
\end{equation}
Therefore, for $\mu=\mu_0(\alpha)$, the first derivative of the dispersion relation near $k=0$ is
\begin{equation}\nonumber
\begin{split}
  \varepsilon'_{0}(\alpha)\sim\frac{Ak^{2\alpha-3}+Bk^{\alpha}+Ck+Dk^{3}}{\sqrt{ak^{2\alpha-2}+bk^{\alpha+1}+ck^{2}+dk^{4}}}.
\end{split}
\end{equation}
When $1<\alpha<2$, one can see that, $\varepsilon'_{0}(\alpha)\sim k^{\alpha-2}$ which is divergent for $k\rightarrow0$.
This implies that along the critical lines, LR hopping will break conformal invariance for $1<\alpha<2$.

\section{The Case of $\alpha=0$}
\begin{figure}
 \centering
\includegraphics[width=.9\linewidth]{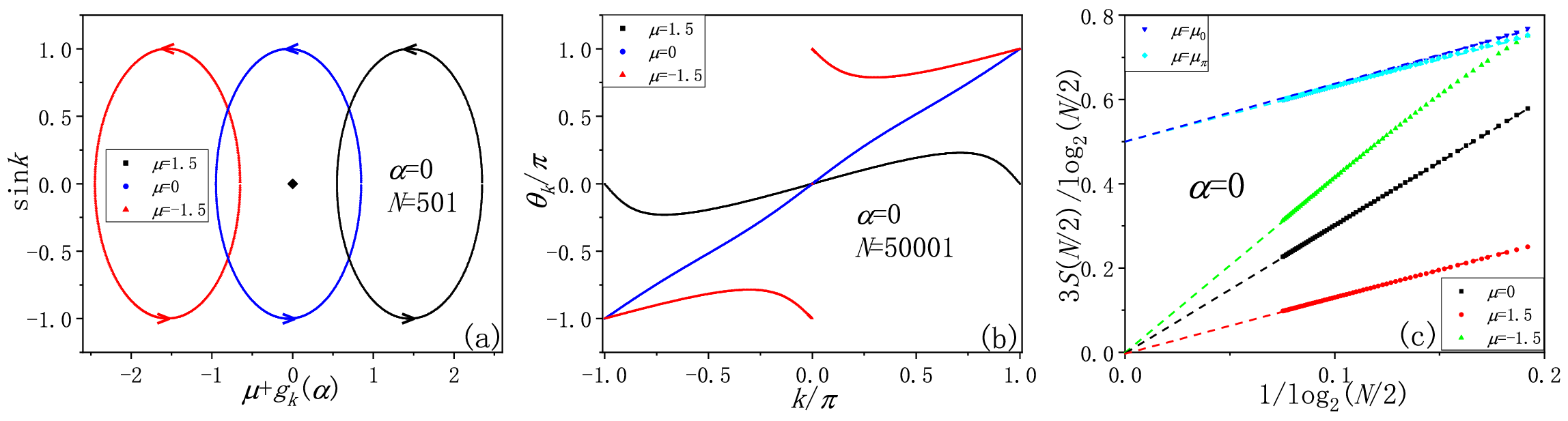}
 \caption{\label{a0} (Color online) (a) The winding vector in the $y-z$ plane.
 (b) The Bogoliubov angle $\theta_k$ as a function of $k$.
 (c) The $3S(N/2)/\log_2(N/2)$ as a function of $1/\log [N/2]$ for various $\mu$ at $\alpha=0$.}
\end{figure}

\subsection{Winding number}
Similar to that in the case of $\alpha>0$, the winding vector and the corresponding Bogoliubov angle $\theta_k$ for $\alpha=0$ are shown in Fig. \ref{a0}(a)-(b), respectively.
However, when $k=0$, the $z$-component of the winding vector $\mu+g_{k=0}(0)$ tends to positive infinity for $N\rightarrow+\infty$.
It means that the $\theta_k$ is singular at $k=0$.
From the theory of the distribution \cite{Bar}, the Eq. (\ref{w}) becomes
\begin{equation}\nonumber
\begin{split}
  \omega\equiv\frac{1}{2\pi}\int_{-\pi}^{\pi} {\rm d}\theta_k=\frac{1}{2\pi}[\int_{-\pi}^{-2\pi/N} {\rm d}\theta_k+\int_{2\pi/N}^{\pi} {\rm d}\theta_k+(\theta_0-\theta_{-2\pi/N})+(\theta_{2\pi/N}-\theta_{0})].
\end{split}
\end{equation}
Apparently, both the Bogoliubov angles $\theta_{k=\pm2\pi/N}$ tend to 0 or $\pi$ for $N\rightarrow+\infty$, so that $(\theta_0-\theta_{-2\pi/N})+(\theta_{2\pi/N}-\theta_{0})\rightarrow0$.
Then the winding number holds
\begin{equation}\nonumber
\begin{split}
  \omega=\frac{1}{2\pi}(\int_{-\pi}^{0^-} {\rm d}\theta_k+\int_{0^+}^{\pi} {\rm d}\theta_k).
\end{split}
\end{equation}

From Fig. \ref{a0}(a)-(b), the phases are topologically trivial for $\mu<\mu_0(0)$ or $\mu>\mu_\pi(0)$ while topologically nontrivial for $\mu_0(0)<\mu<\mu_\pi(0)$ .
In other words, the entropy and topological properties for $\alpha=0$ are similar to those in the case of $\alpha>2$.

\subsection{von Neumann entropy}
From Fig. 1(e) in the letter, there are two QPT points at $\mu=\mu_0(\alpha)$ and $\mu_\pi(\alpha)$ for $\alpha=0$.
Similarly, Fig. \ref{a0}(c) shows the $3S(N/2)/\log_2(N/2)$ with respect to $1/\log_2(N/2)$ for various $\alpha$ and $\mu$.
From the figure we can find that the effective central charge $c_{{\rm eff}}=0.5$ at the two QPT points while $c_{{\rm eff}}=0$ in the gapped region.

\section{The Case of $-2<\alpha<0$}

\begin{figure}
 \centering
\includegraphics[width=.9\linewidth]{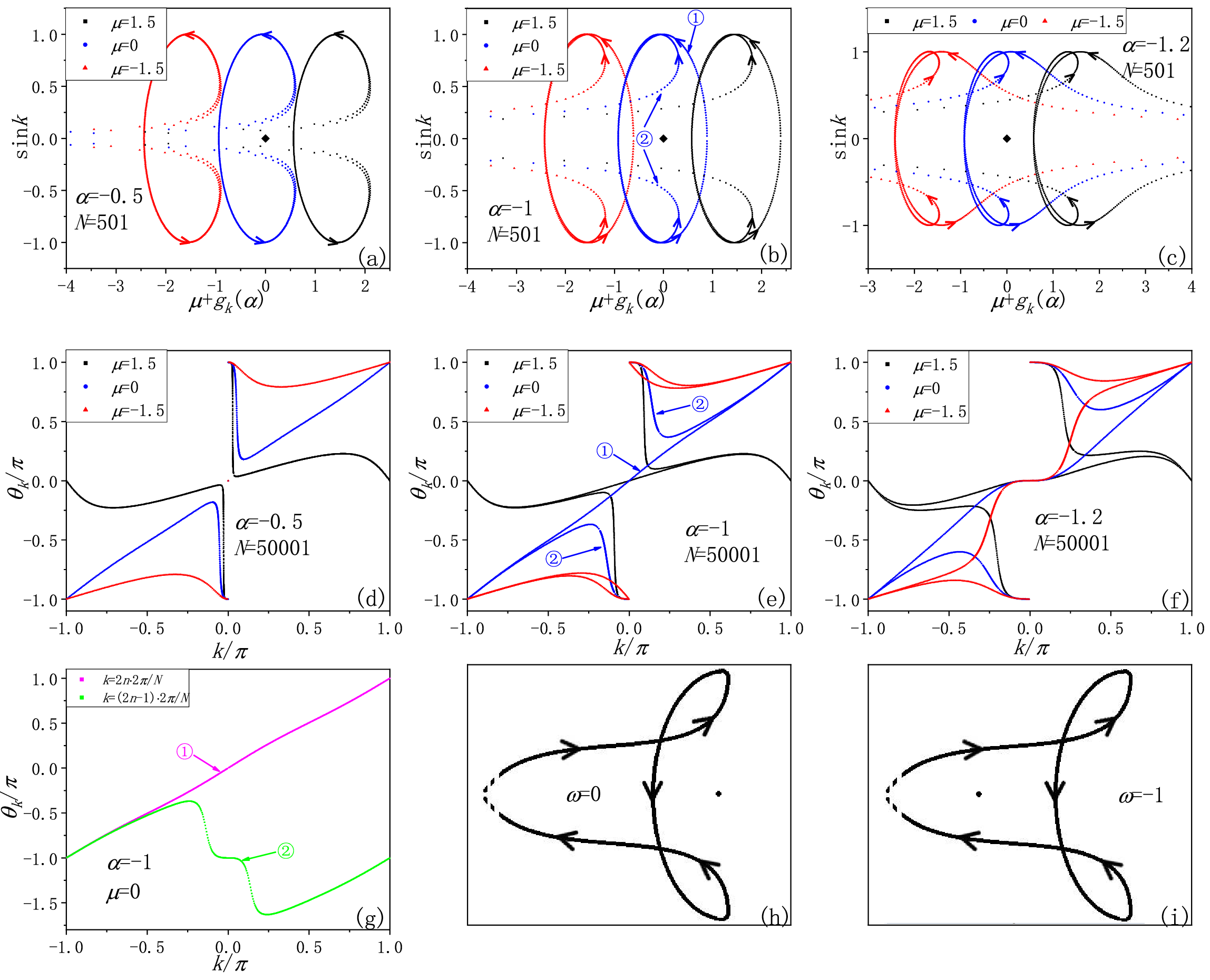}
 \caption{\label{al} (Color online) (a)-(c) The winding vector in the $y-z$ plane for $\alpha=-0.5$, $-1$, and $-1.2$, respectively.
 (d)-(f) The Bogoliubov angle $\theta_k$ as functions of $k$ for $\alpha=-0.5$, $-1$, and $-1.2$, respectively.
 (g) The Bogoliubov angle $\theta_k$ as a function of $k$ for $\alpha=-1$ and $\mu=0$.
 (h)-(i) The schematic diagram of the loop of winding vector.}
\end{figure}

\subsection{Winding number}
Similarly, the winding vector and the Bogoliubov angle $\theta_k$ for $\alpha=-0.5$, $-1$, and $-1.2$ are shown in Fig. \ref{al}(a)-(f), respectively.
Just as the Lee-Yang zeros and QES mentioned above, the Bogoliubov angle $\theta_k$ is discontinuous at all $k$.
Then, from the theory of the distribution \cite{Bar}, the Eq. (\ref{w}) becomes
\begin{equation}\label{ent}
\begin{split}
  \omega\equiv\frac{1}{2\pi}\int_{-\pi}^{+\pi} {\rm d}\theta_k=\frac{1}{2\pi}\sum_{i=1}^{N-1}(\theta_{k_{i+1}}-\theta_{k_i}),
\end{split}
\end{equation}
where $k_{i}=(2i-1-N)\pi/N$ and $N\rightarrow+\infty$.

From Fig. \ref{al}(a)-(f), it can be found that the loop of winding vector and Bogoliubov angle are divided into two branches.
The two branches correspond to $k=2n\cdot2\pi/N$ and $k=(2n-1)\cdot2\pi/N$, respectively.
For finite $N$, the $k_1=-[N/2]\cdot2\pi/N$ and $k_{N}=[N/2]\cdot2\pi/N$.
Therefore, the summation in the Eq. (\ref{ent}) is dependent on the parity of $[N/2]$.
In order to obtain the winding number, we consider a typical case of $\mu=0$ and $\alpha=-1$.
The corresponding winding vector and Bogoliubov angle are marked with blue points in Fig. \ref{al}(b) and (e).
The two branches are marked with \ding{172} (corresponding to $k=2n\cdot2\pi/N$) and \ding{173} [corresponding to $k=(2n-1)\cdot2\pi/N$], respectively.
For even $[N/2]$, both $\theta_{k_1}$ and $\theta_{k_N}$ are on the branch \ding{172} and the loop of branch \ding{172} in Fig. \ref{al}(b) winds around the origin counterclockwise which indicates the winding number $\omega=+1$ [also see the pink curve in Fig.\ref{al}(g)].
While for odd $[N/2]$, both $\theta_{k_1}$ and $\theta_{k_N}$ are on the branch \ding{173} and the loop of branch \ding{173} in Fig. \ref{al}(b) doesn't wind around the origin [see the schematic Fig. \ref{al}(h)]. Hence, the winding number $\omega=0$ [see the green curve in Fig.\ref{al}(g)].
Therefore, the summation in the Eq. (\ref{ent}) is similar to those in the case of the $\sum_{m=1}^{N}(-1)^{m-1}$.
The $\sum_{m=1}^{N}(-1)^{m-1}$ equals to 0 (1) for even (odd) $N$.
When $N\rightarrow+\infty$, $\sum_{m=1}^{\infty}(-1)^{m-1}:=\eta(0)$ and $\eta(0)=1/2$ by means of analytic continuation \cite{Edw}.
Accordingly, we define that the winding number $\omega:=+1/2$.

Correspondingly, we study the winding numbers of $-1<\alpha<0$, $\alpha=-1$, and $-2<\alpha<-1$, respectively.
When $-1<\alpha<0$, there is only one QPT point at $\mu=\mu_\pi(\alpha)$. In Fig. \ref{al}(a) and (d),
we plot the winding vector and the corresponding Bogoliubov angle for $\alpha=-0.5$.
From them it is found that the winding vector and the Bogoliubov angle of the two branches are similar.
The loop of winding vector for $\mu<\mu_\pi(\alpha)$ doesn't winds around the origin in the $y-z$ plane.
Therefore, the winding numbers $\omega=0$ in this region.
However, the loop of winding vector for $\mu>\mu_\pi(\alpha)$ winds around the origin clockwise [see the schematic Fig. \ref{al}(i)]
and the winding numbers $\omega=-1$ in this region.
So the phase is topologically (non)trivial for $\mu<(>)\mu_\pi(\alpha)$.

At the $\alpha=-1$, there are two QPT points at $\mu=\mu_0(-1)$ and $\mu_\pi(-1)$. Similarly, we show
the winding vector and corresponding Bogoliubov angle $\theta_k$ in Fig. \ref{al}(b) and (e).
For $\mu<\mu_0(0)$ both loops of winding vector don't wind around the origin [see red curves in Fig. \ref{al}(b)] and the winding number $\omega=0$.
The situation of $\mu_0(-1)<\mu<\mu_\pi(-1)$ is similar to that in the above discussion and the winding number $\omega:=+1/2$.
For $\mu>\mu_\pi(\alpha)$  the $\omega$ is 0 for one branch and $-1$ for the another branch [see the schematic Fig. \ref{al}(i)].
So the winding number $\omega:=-1/2$.
Conclusively, the phase is topologically trivial for $\mu<\mu_0(-1)$.
The phases are topologically nontrivial for $\mu_0(-1)<\mu<\mu_\pi(-1)$ with $\omega:=+1/2$ and for $\mu>\mu_\pi(-1)$ with $\omega:=-1/2$.

For $-2<\alpha<-1$, there is only one QPT point at $\mu=\mu_\pi(\alpha)$.
Similarly, the winding vector and the Bogoliubov angle of $\alpha=-1.2$ are shown in Fig. \ref{al}(c) and (f).
We can find that the winding number $\omega:=+1/2(-1/2)$ for $\mu<(>)\mu_\pi(\alpha)$.

\begin{figure}
 \centering
\includegraphics[width=.9\linewidth]{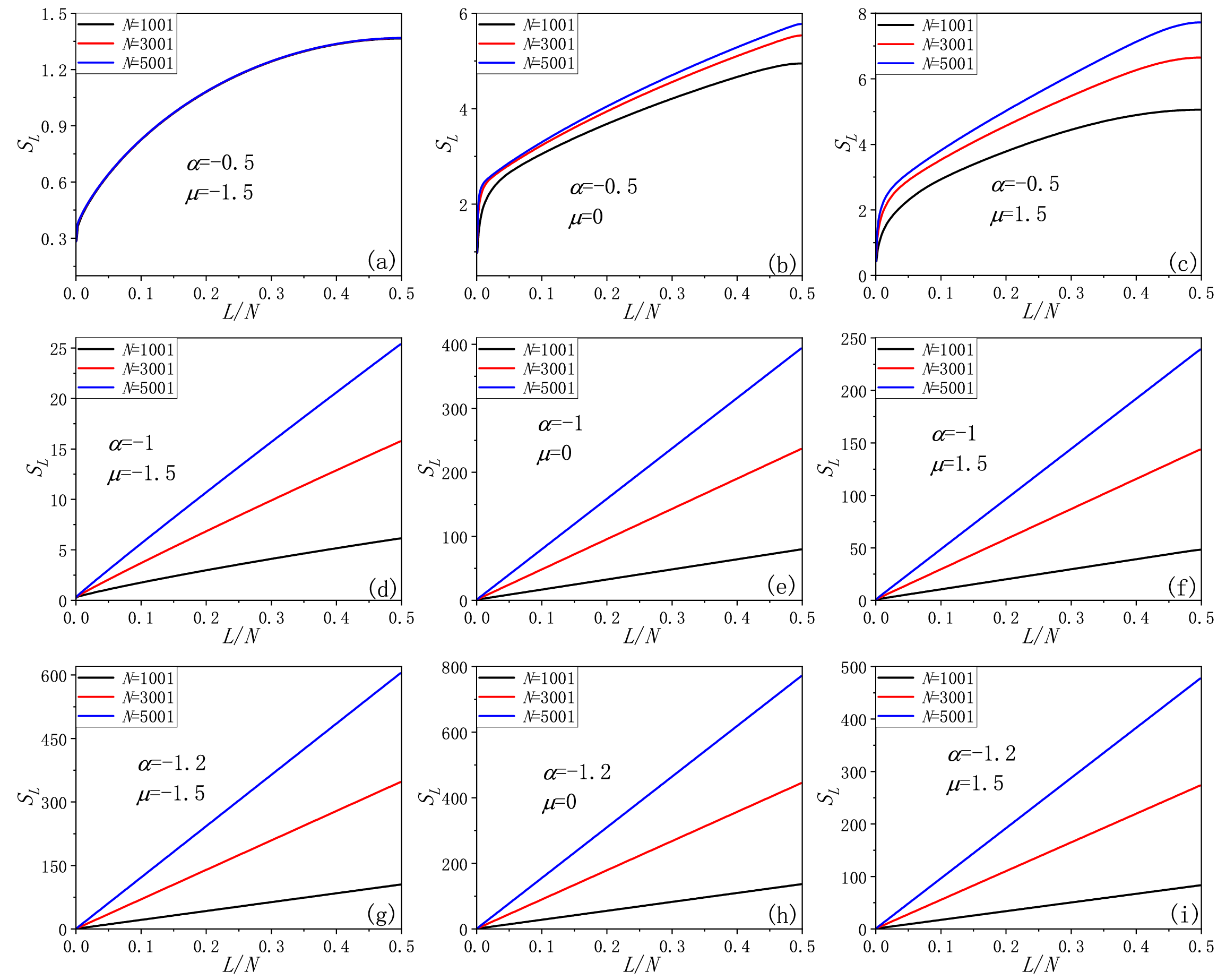}
 \caption{\label{s2} (Color online) The von Neumann entropy $S_L$ as functions of the $L/N$ for various $\alpha$ and $\mu$.
 The black, red, and red lines represent to $N=1001$, $3001$, and $5001$, respectively.}
\end{figure}

\begin{figure}
 \centering
\includegraphics[width=.9\linewidth]{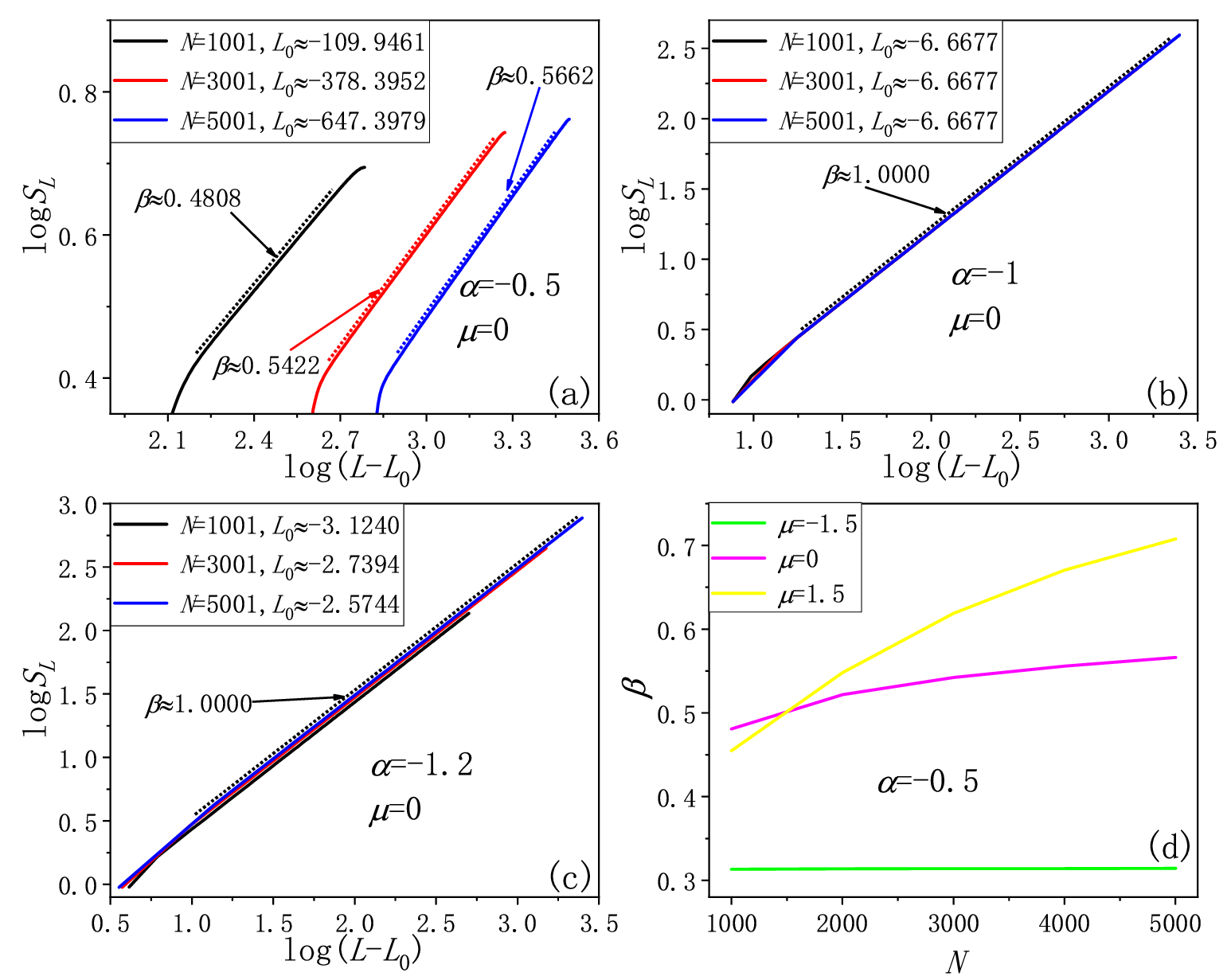}
 \caption{\label{s4} (Color online) (a)-(c) The logarithm of von Neumann entropy $\log S_L$ as functions of $\log(L-L_0)$ for various $\alpha$ with $\mu=0$.
 Here, the black, red, and red lines represent to $N=1001$, $3001$, and $5001$, respectively. (d) The exponent $\beta$ as functions of the system size $N$ for $\alpha=-0.5$.
 Here, the green, pink, and yellow lines represent to $\mu=-1.5$, $0$, and $1.5$, respectively.}
\end{figure}

\begin{figure}
 \centering
\includegraphics[width=.9\linewidth]{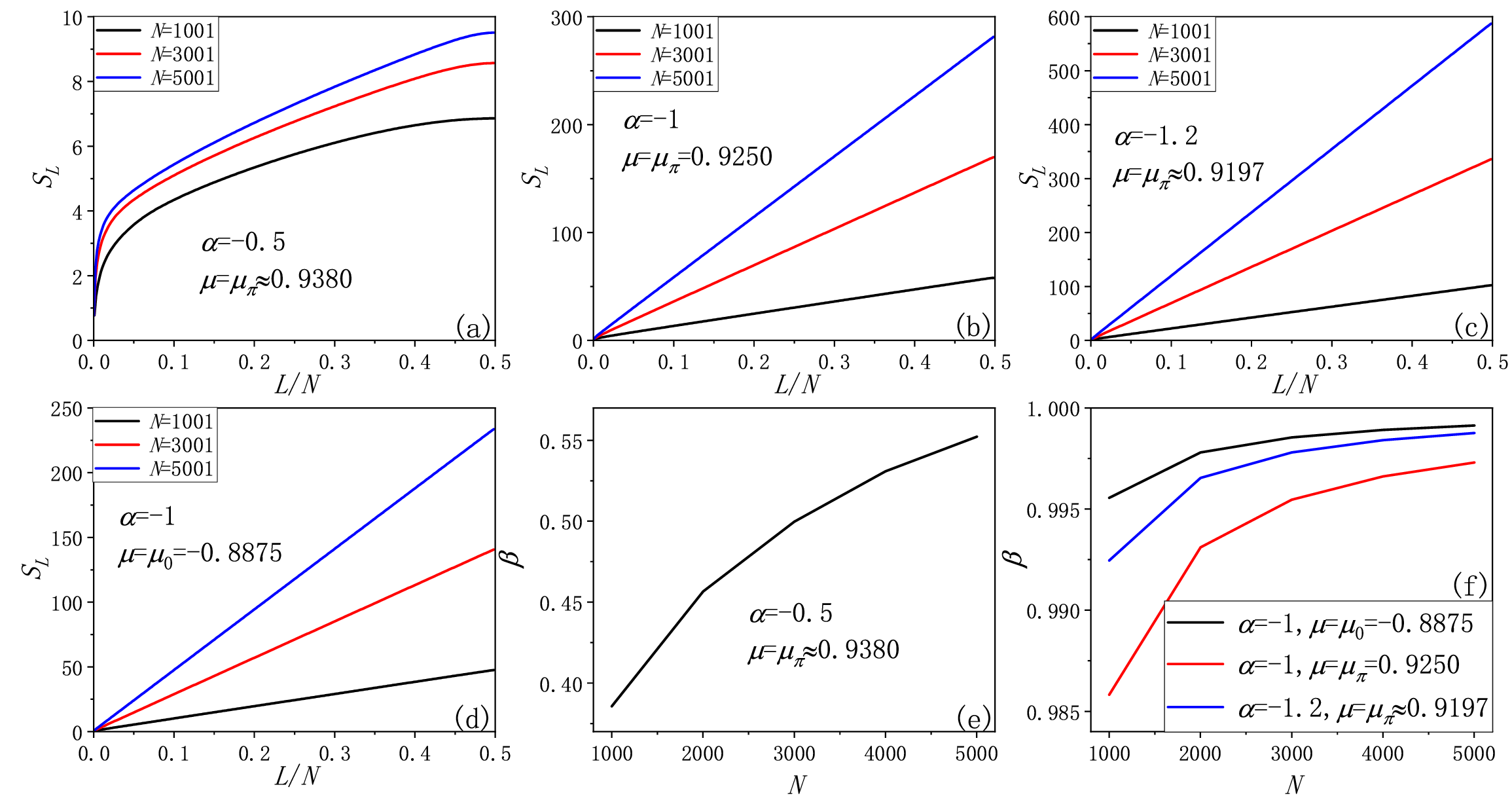}
 \caption{\label{s3} (Color online) (a)-(d) The von Neumann entropy $S_L$ as functions of the $L/N$ at the QPT points $\mu_0(\alpha)$ or $\mu_\pi(\alpha)$ for $\alpha=-0.5$, $-1$, and $-1.2$, respectively.
 (e)-(f) The corresponding exponent $\beta$ as functions of the system size $N$.}
\end{figure}

\subsection{von Neumann entropy}
In this subsection, we discuss the von Neumann entropy $S_L$ for $-2<\alpha<0$.
The $S_L$ as functions of the relative size of subchain $L/N$ are shown in Fig. \ref{s2} for various $\alpha$ and $\mu$ in the gapped region.
The black, red, and blue lines correspond to $N=1001$, $3001$, and $5001$, respectively.
We can find that the relationship between $S_L$ and $L$ is well fitted by $S(L)=A(L-L_0)^{\beta}$ with nonuniversal parameter $A$ and $L_0$.
In Fig. \ref{s4}(a)-(c), we also plot the $\log S_L$ as functions of $\log(L-L_0)$ for various $\alpha$ at $\mu=0$.
It can be seen that the $\beta\approx 1$ for $-2<\alpha\le -1$ and $\beta$ is dependent on the $N$ for $-1<\alpha<0$. When $N$ increases, $\beta$ goes to finite values between $0$ and $1$ [see Fig. \ref{s4}(d)]
.

At the QPT points of $-2<\alpha<0$, the situations are similar.  Fig. \ref{s3}(a)-(d) show the $S_L$ as functions of $L/N$ at the QPT points for various $\alpha$ and the corresponding exponent $\beta$ as functions of the system size $N$ are also shown in Fig. \ref{s3}(e)-(f).

Summarily, whether at QPT points or in the gapped region the von Neumann entropy of subchain $S(L)=A(L-L_0)^{\beta}$ and the exponent $0<\beta<1$ and $\beta\approx1$ for $\alpha>-1$ and $-2<\alpha\le-1$, respectively.

\section{The Special QPT Points $\mu_0(0)$ and $\mu_0(-1)$}
\begin{figure}
 \centering
\includegraphics[width=.9\linewidth]{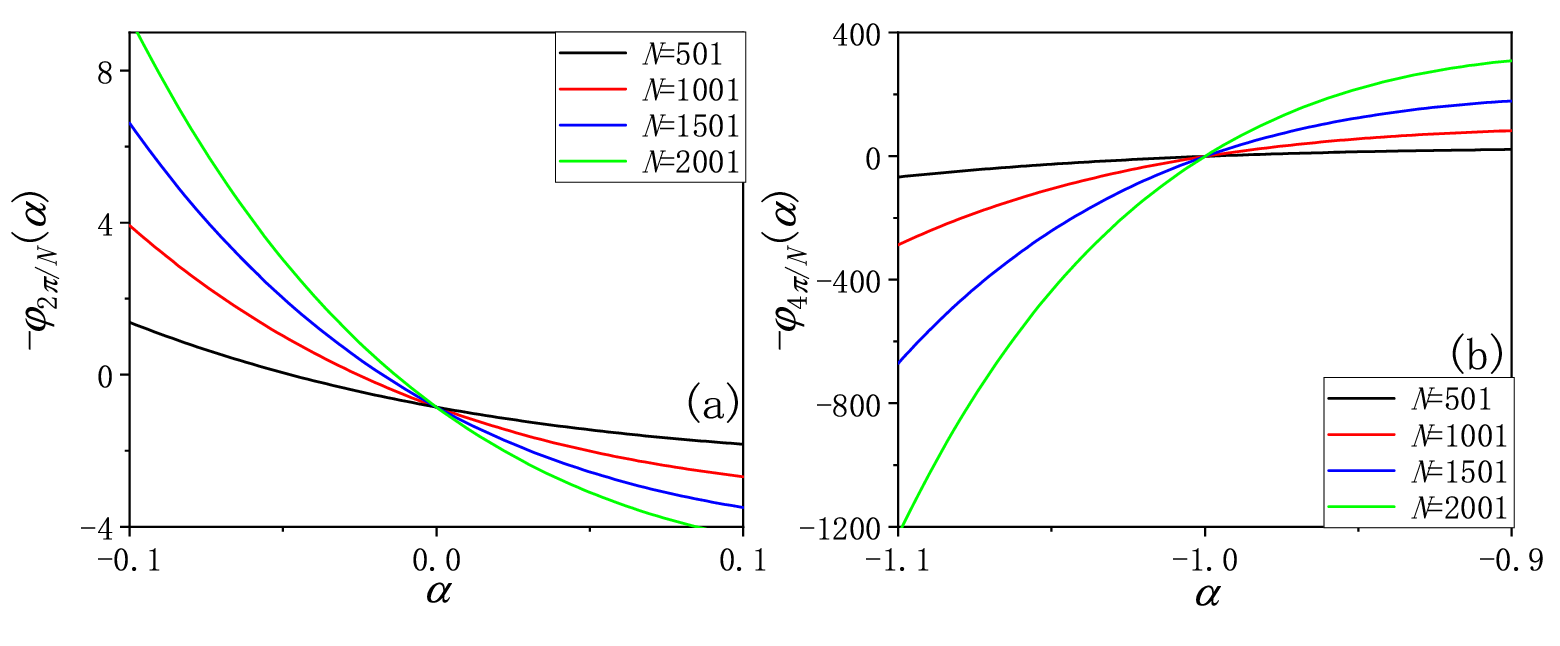}
 \caption{\label{gk} (Color online) (a) $-\phi_{2\pi/N}(\alpha)$ and (b) $-\phi_{4\pi/N}(\alpha)$ as functions of $\alpha$.
 The black, red, blue, and green lines correspond to $N=501$, 1001, 1501, and 2001, respectively.
 (c) $\phi_{\pi-\pi/N}(\alpha)$ as a function of $\alpha$.
 The black line represents the analytic continuation value of the Dirichlet eta function.
 The red, blue, green, and grey lines correspond to $N=101$, 1001, 10001, and 100001, respectively.}
\end{figure}
From the Fig. 1 and 2 in the letter, we can find that there are two isolated QPT points at $\mu=\mu_0(\alpha=0)$ and $\mu_0(\alpha=-1)$.
According to the distribution of Lee-Yang zeros, it is found that the zeros corresponding to $k=0$ tend to $-\infty$ for $\alpha=0$ and $-1$, respectively.
However, the zeros corresponding to $k=\pm (2n+1)2\pi/N$ goes to $\infty$ for $\alpha=-1$. The other zeros near the $k=0$ will converge to real axis when $N\rightarrow+\infty$. Therefore, the zeros closest to the real axis (ZCRAs) correspond to $k=\pm2\pi/N$ ($\pm4\pi/N$) for $\alpha=0$ ($-1$).
Due to the equation
\begin{equation}
\begin{split}\label{muQPT}
  \mu(k,\alpha)=-(1-f)\cos k-f\phi_{k}(\alpha),
\end{split}
\end{equation}
we discuss the values of $\phi_{\pm2\pi/N}(0)$ and $\phi_{\pm4\pi/N}(-1)$ in the following.

Now, we consider $\phi_k(0)$, which yields
\begin{equation}
\begin{split}\label{phi4}
  \phi_k(0)=\sum_{m=1}^{[N/2]}\cos(mk)=\frac{\sin(Nk/2)-\sin(k/2)}{2\sin(k/2)}.
\end{split}
\end{equation}
Considering $k=\pm2\pi/N$, the Eq. (\ref{phi4}) can be transformed into
\begin{equation}\nonumber
\begin{split}
  \phi_{\pm2\pi/N}(0)=\frac{\sin\pi-\sin(\pi/N)}{2\sin(\pi/N)}=-\frac{1}{2}.
\end{split}
\end{equation}
Therefore, for $N\rightarrow+\infty$, the value of the QPT point $\mu_0(0)=-(1-f)\cos\frac{2\pi}{N}-f\phi_{2\pi/N}(0)\rightarrow-1+3f/2$.

When $\alpha=-1$, we can obtain $\phi_{\pm4\pi/N}(-1)$ from
the Eq. (\ref{phi3}), which yields
\begin{equation}\nonumber
\begin{split}
  \phi_{\pm4\pi/N}(-1)=\frac{\cos(2\pi/N)-1}{4\sin^2(2\pi/N)}\stackrel{N\rightarrow+\infty}{\longrightarrow}-\frac{1}{8}.
\end{split}
\end{equation}
Therefore, for $N\rightarrow+\infty$, the value of the QPT point $\mu_0(-1)=-(1-f)\cos\frac{4\pi}{N}-f\phi_{4\pi/N}(-1)\rightarrow-1+9f/8$.
For $f=0.1$, the $\mu_0(0)=-0.85$ and $\mu_0(-1)=-0.8875$ which is the same as that in the letter.

Finally, in order to understand why there are only two isolated QPT points at $\mu=\mu_0(0)$ and $\mu_0(-1)$, we plot the $-\phi_{2\pi/N}(\alpha)$ and $-\phi_{4\pi/N}(\alpha)$ as functions of $\alpha$ in Fig. \ref{gk}(a)-(b), respectively.
The black, red, blue, and green lines correspond to $N=501$, 1001, 1501, and 2001, respectively.
We can find that $-\phi_{2\pi/N}(\alpha)$ tends to $-\infty$ ($+\infty$) for $\alpha>(<)0$ as $N\rightarrow+\infty$.
Only at $\alpha=0$, $-\phi_{2\pi/N}(\alpha)$ converges to a finite value as $N\rightarrow+\infty$.
Likewise, $-\phi_{4\pi/N}(\alpha)$ tends to $+\infty$ ($-\infty$) for $\alpha>(<)-1$ as $N\rightarrow+\infty$.
Only at $\alpha=-1$, $-\phi_{4\pi/N}(\alpha)$ converges to a finite value as $N\rightarrow+\infty$.
Therefore, there are QPT points at $\mu=\mu_0(\alpha)$ only for $\alpha=0$ and $-1$.

\section{The Case of $\alpha\le-2$}
\subsection{Quantum phase transitions}
\begin{figure}
 \centering
\includegraphics[width=.9\linewidth]{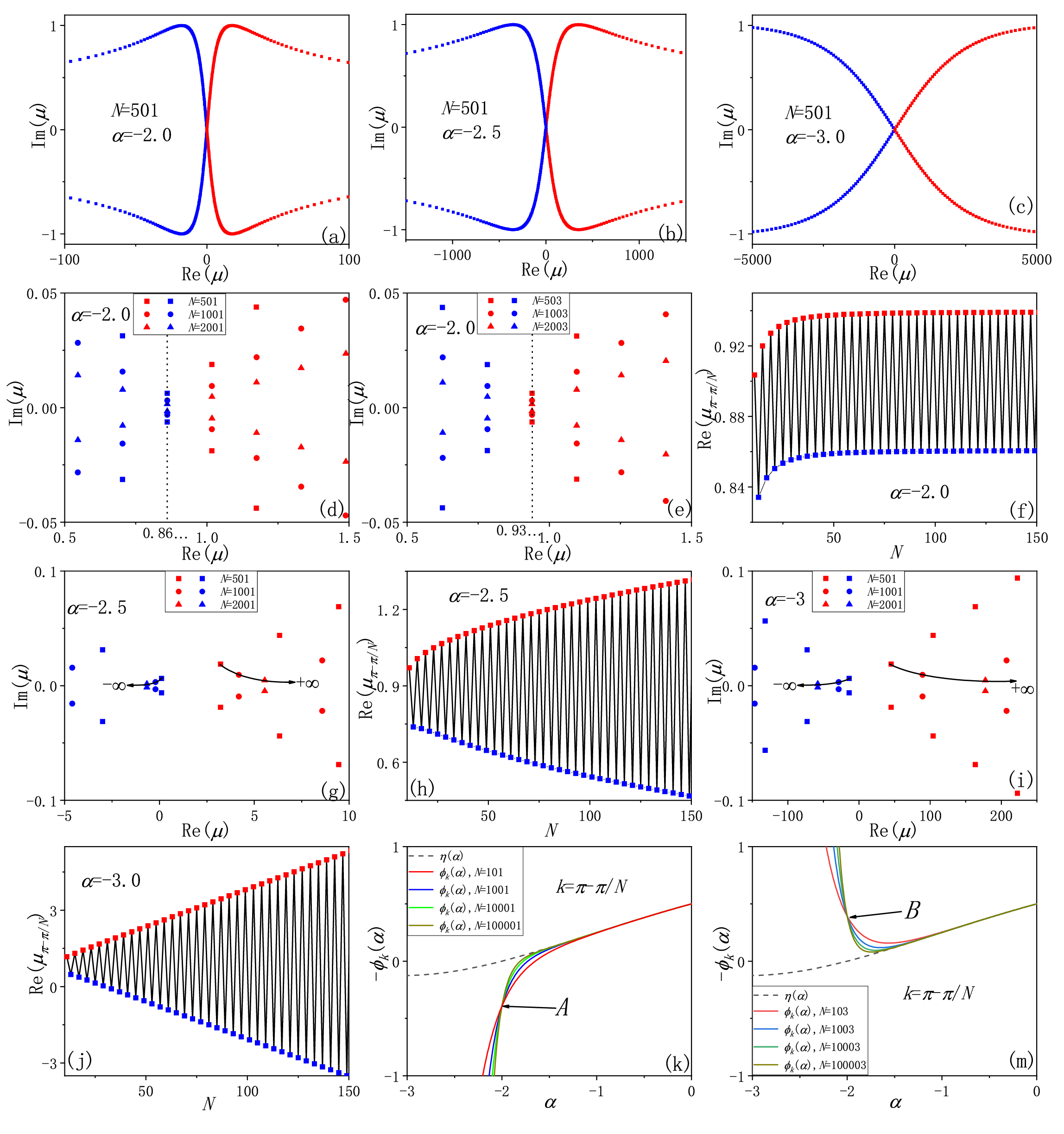}
 \caption{\label{aln2} (Color online) (a), (b), and (c) are the Lee-Yang zeros of $\varepsilon_k=0$ in the complex $\mu$ plane for $\alpha=-2$, $-2.5$ and $-3$, respectively.
 (d) and (e) are the zeros of $\alpha=-2$ for even and odd $[N/2]$, respectively.
 (f) is the real part of the zeros closest to the real axis (ZCRAs) $\mu_{\pi-\pi/N}$ as a function of $N$.
 (g) and (h) are the zeros of $\alpha=-2.5$ and $-3$ for even $[N/2]$, respectively.
 The block arrow indicates the asymptotic behavior of the ZCRAs.
 (i) and (j) are the real parts of the ZCRAs $\mu_{\pi-\pi/N}$ as functions of $N$ for $\alpha=-2.5$ and $3$, respectively.
 (k) and (m) are the $-\phi_{\pi-\pi/N}(\alpha)$ as functions of $\alpha$ with different $N$ for even and odd $[N/2]$, respectively.
 }
\end{figure}

\par The three typical Lee-Yang zeros for $\alpha=-2$, $-2.5$, and $-3$ with $N=501$ are shown in Fig. \ref{aln2}(a), (b), and (c), respectively.
The blue and red dots correspond to $k=2n\cdot2\pi/N$ and $k=(2n-1)\cdot2\pi/N$, respectively.
Similar to that in the case of $-2<\alpha<0$, we can find that the zeros are divided into two branches and the red (blue) zeros goes to $+\infty$ ($-\infty$) as $k\rightarrow 0$. At same time, both of the two branches for $k\rightarrow\pi$ converge to the same point on the real axis.
However, the value of this point is dependent on the parity of $[N/2]$ for $N\rightarrow \infty$. This is due to the ZCRAs depend on the parity of $[N/2]$ for finite $N$ and quite different from that in the case $-2<\alpha<0$. Furthermore, the asymptotic behaviors of the ZCRAs for $\alpha=-2$ is different from that of $\alpha<-2$. In the following, we shall give detailed discussions of zeros for $\alpha=-2$ and $\alpha<-2$.

For $\alpha=-2$, we plot the zeros for different $N$ with even and odd $[N/2]$ in Fig. \ref{aln2}(d) and (e), respectively.
For a finite $N$, the ZCRAs correspond to $k=\pm(\pi-\pi/N)$.
For even (odd) $[N/2]$,  the ZCRAs are on the red (blue) branch and goes to $0.8607\cdots$ ($0.9393\cdots$) as $N\rightarrow \infty$.
In order to see this phenomenon more clearly, we show the real parts of the zeros $\mu_{\pi-\pi/N}$ as function of $N$ in Fig. \ref{aln2}(f).
Therefore, as $N\rightarrow \infty$, the limitation of ZCRAs is divergent.
This means there is no QPT point for $\alpha=-2$.

In order to unambiguously understand the above properties, we analytically study $\phi_{\pm(\pi-\pi/N)}(\alpha=-2)$.
By considering the Eq. (\ref{phi4}), the Eq. (\ref{phi}) can be expressed as
\begin{equation}\nonumber
\begin{split}
  \phi_{k}(-2)&=\sum_{m=1}^{[N/2]}m^2\cos(mk)=-\frac{{\rm d^2}}{{\rm d}k^2}\sum_{m=1}^{[N/2]}\cos(mk)\\
  &=\frac{(N^2-1)\sin(Nk/2)}{8\sin(k/2)}+\frac{N\cos(k/2)\cos(Nk/2)}{4\sin^2(k/2)}+\frac{\cos(k/2)\sin(Nk/2)}{4\sin^3(k/2)}.
\end{split}
\end{equation}
When $k=\frac{2\pi}{N}h$,
\begin{equation}\nonumber
\begin{split}
  \phi_{k}(-2)\equiv\phi_{h}(-2)=\frac{N\cos(h\pi/N)\cos(h\pi)}{4\sin^2(h\pi/N)}.
\end{split}
\end{equation}
For $h=\frac{N-1}{2}$ and $N\rightarrow+\infty$,
\begin{equation}\nonumber
 \phi_{h}(-2)\rightarrow\frac{\pi\cos(h\pi)}{8}=\left\{\begin{array}{cc} 0.3926\cdots,\qquad {\rm for\; even}\; h,\qquad\\
 [1.5mm] -0.3926\cdots,\qquad {\rm for\; odd}\; h.\qquad \end{array}\right.
\end{equation}
From Eq. (\ref{muQPT}), we obtain $\mu_{\pm(\pi-\pi/N)}=0.8607\cdots$ and $0.9393\cdots$ for even and odd $h$ when $f=0.1$, which are equal to the numerical results in Fig. \ref{aln2}(d)-(f).

The asymptotic behaviors of the ZCRAs for $\alpha<-2$ are quite different from that of $\alpha=-2$. When $N\rightarrow\infty$, the real parts of each ZCRAs goes to $-\infty$ ($+\infty)$ for even (odd) $[N/2]$.  As two typical cases of $\alpha=-2.5$ and $-3$, the zeros for even $[N/2]$ are shown in Fig. \ref{aln2}(g) and (h), respectively.
As indicated by the black arrow in the figures, we can find that the real parts of the ZCRAs are divergent to $\pm\infty$.
Similarly, we plot the real parts of the ZCRAs $\mu_{\pi-\pi/N}$ as functions of $N$ in Fig. \ref{aln2}(i) and (j), respectively.
This means that the real parts of the ZCRAs are divergent in the thermodynamic limit ($N\rightarrow\infty$).
Therefore, no QPT point occurs for $\alpha<-2$.
Additionally, the case where $[N/2]$ is odd can be analyzed analogously to the even case.

Similarly, we can obtain the Eq. (\ref{phi}) analytically at $\alpha=-3$. That is
\begin{equation}\nonumber
\begin{split}
  \phi_{k}(-3)=&\sum_{m=1}^{[N/2]}m^3\cos(mk)=-\frac{{\rm d^3}}{{\rm d}k^3}\sum_{m=1}^{[N/2]}\sin(mk)\\
  =&\frac{N(N^2-1)\sin(Nk/2)}{16\sin(k/2)}+\frac{3(N^2-1)\cos(k/2)\cos(Nk/2)}{16\sin^2(k/2)}-\\
  &\frac{[2\cos^2(k/2)+1][N\sin(k/2)\sin(Nk/2)+\cos(k/2)\cos(Nk/2)-1]}{8\sin^4(k/2)}.
\end{split}
\end{equation}
When $k=\frac{2\pi}{N}h$,
\begin{equation}\nonumber
\begin{split}
  \phi_{k}(-3)\equiv\phi_{h}(-3)=\frac{3(N^2-1)\cos(h\pi/N)\cos(h\pi)}{16\sin^2(h\pi/N)}-\frac{[2\cos^2(h\pi/N)+1][\cos(h\pi/N)\cos(h\pi)-1]}{8\sin^4(h\pi/N)}.
\end{split}
\end{equation}
For $h=\frac{N-1}{2}$ and $N\rightarrow+\infty$,
\begin{equation}\nonumber
 \phi_{h}(-3)\rightarrow\frac{3\pi N\cos(h\pi)}{32}\rightarrow\left\{\begin{array}{cc} +\infty,\qquad {\rm for\; even}\; h,\qquad\\
 [1.5mm] -\infty,\qquad {\rm for\; odd}\; h.\qquad \end{array}\right.
\end{equation}

To systematically analyze the absence of QPT points for $\alpha\le-2$, we plot $-\phi_{\pi-\pi/N}(\alpha)$ as functions of $\alpha$ across different $N$ in Fig. \ref{aln2}(k) and (m). The two figures are categorized by the parity of $[N/2]$, with Fig. \ref{aln2}(k) and (m) corresponding to even and odd values, respectively.
We can find that $-\phi_{\pi-\pi/N}(\alpha)$ tends to $\eta(\alpha)$ for $\alpha>-2$ regardless of the parity of $[N/2]$. However, when $\alpha<-2$, $-\phi_{\pi-\pi/N}(\alpha)$ tends to $+\infty$ ($-\infty$) for even (odd) $[N/2]$ as $N\rightarrow\infty$.
At $\alpha=-2$, $-\phi_{\pi-\pi/N}(\alpha)$ goes to two finite values $\pm\pi/8$ [points $A$ and $B$ in Fig. \ref{aln2}(k) and (m), respectively], which is consistent with the analysis results.

\subsection{Winding number}

\begin{figure}
 \centering
\includegraphics[width=.9\linewidth]{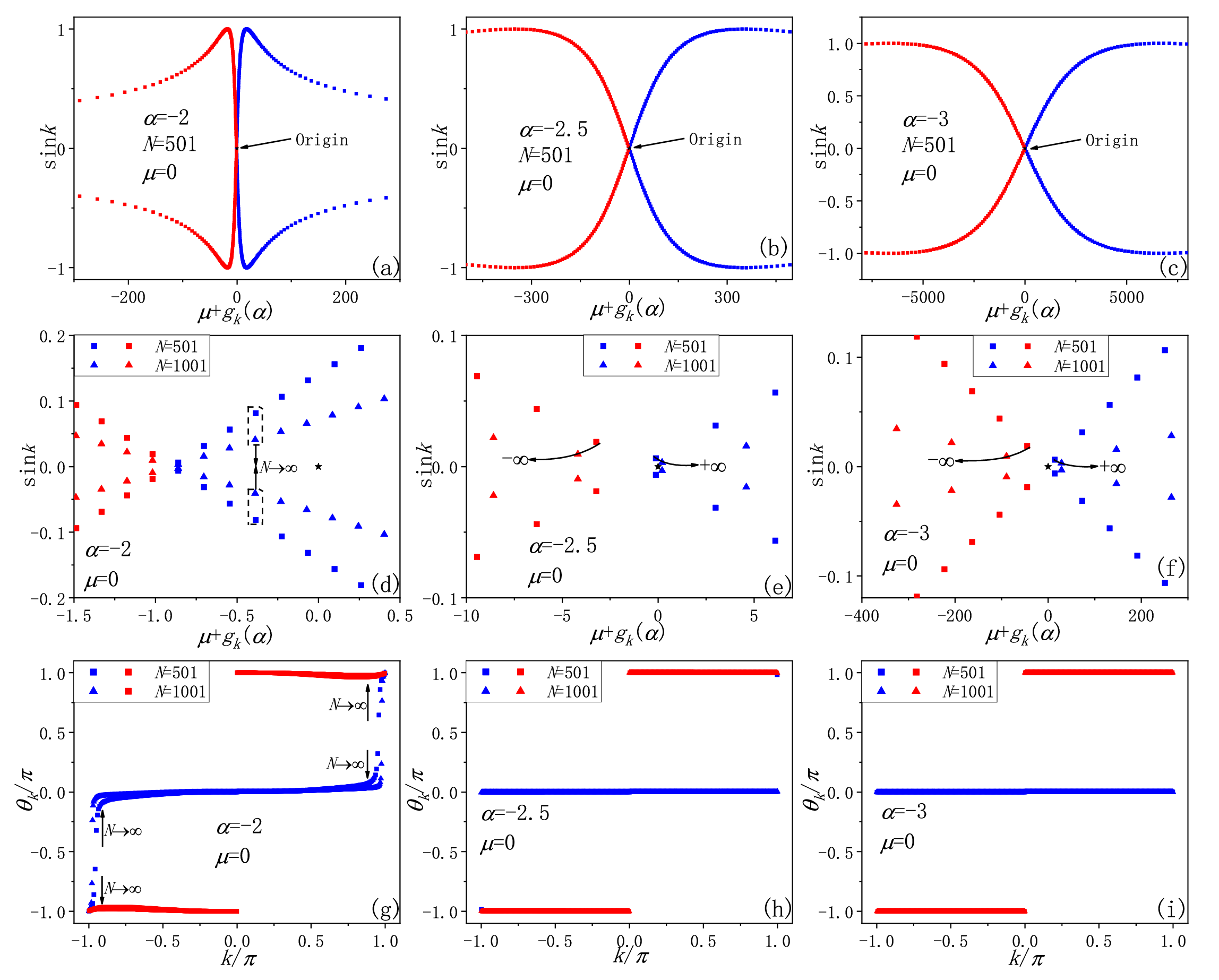}
 \caption{\label{wnln2} (Color online) (a)-(c) The winding vector in the $y-z$ plane for $\alpha=-2$, $-2.5$, and $-3$, respectively.
 (d)-(f) The winding vector near the origin for various $N$ and $\alpha$.
 (g)-(i) The Bogoliubov angle $\theta_k$ as a function of $k$ for $\alpha=-2$, $-2.5$, and $-3$, respectively.
 }
\end{figure}

Similar to that in the case of $-2<\alpha<0$, the winding vectors are given in Fig. \ref{wnln2}(a)-(c) for various $\alpha$ and $\mu$.
In the figures the blue and red dots correspond to $k=2n\cdot2\pi/N$ and $k=(2n-1)\cdot2\pi/N$, respectively.
In order to see the winding number in the $N\rightarrow \infty$, we also plot the winding vectors and Bogoliubov angles for different $N$ in Figs. \ref{wnln2}(d)-(f) and Figs. \ref{wnln2}(g)-(i), respectively.
From these figures, it can be found that the loop of winding vectors and Bogoliubov angles are divided into two branches.
For $\alpha=-2$, as $N$ increases, we can find from Fig. \ref{wnln2}(d) that the winding vectors tend to the axis of $\sin k=0$ and finite blue vectors in the left side of the origin, which is marked by the black pentagram.
It means that the Bogoliubov angles of winding vectors corresponding to $k=(2n+1)\cdot2\pi/N$ always tend to $\pm\pi$.
And the Bogoliubov angles of winding vectors corresponding to $k=2n\cdot2\pi/N$ tend to 0 or $\pm\pi$.
From Figs. \ref{wnln2}(d) and (g), we can obtain that only a finite subset of Bogoliubov angles equal to $\pm\pi$ corresponding to $k=2n\cdot2\pi/N$ as $N\rightarrow\infty$ which does not contribute to the winding number (see V.A in this Supplemental Material).
Therefore, the winding number $\omega=0$ for $\alpha=-2$.

For $\alpha=-2.5$ and $-3$, it is found that the winding vectors are divergent to $\pm\infty$ for $N\rightarrow\infty$ [the directions are shown by the black arrows in Fig. \ref{wnln2}(e)-(f)].
And the contributions of winding number of the two branches are both 0 for any finite $\mu$.
Then, we can conclude that the winding number $\omega=0$ for $\alpha<-2$.

\subsection{von Neumann entropy}
\begin{figure}
 \centering
\includegraphics[width=.9\linewidth]{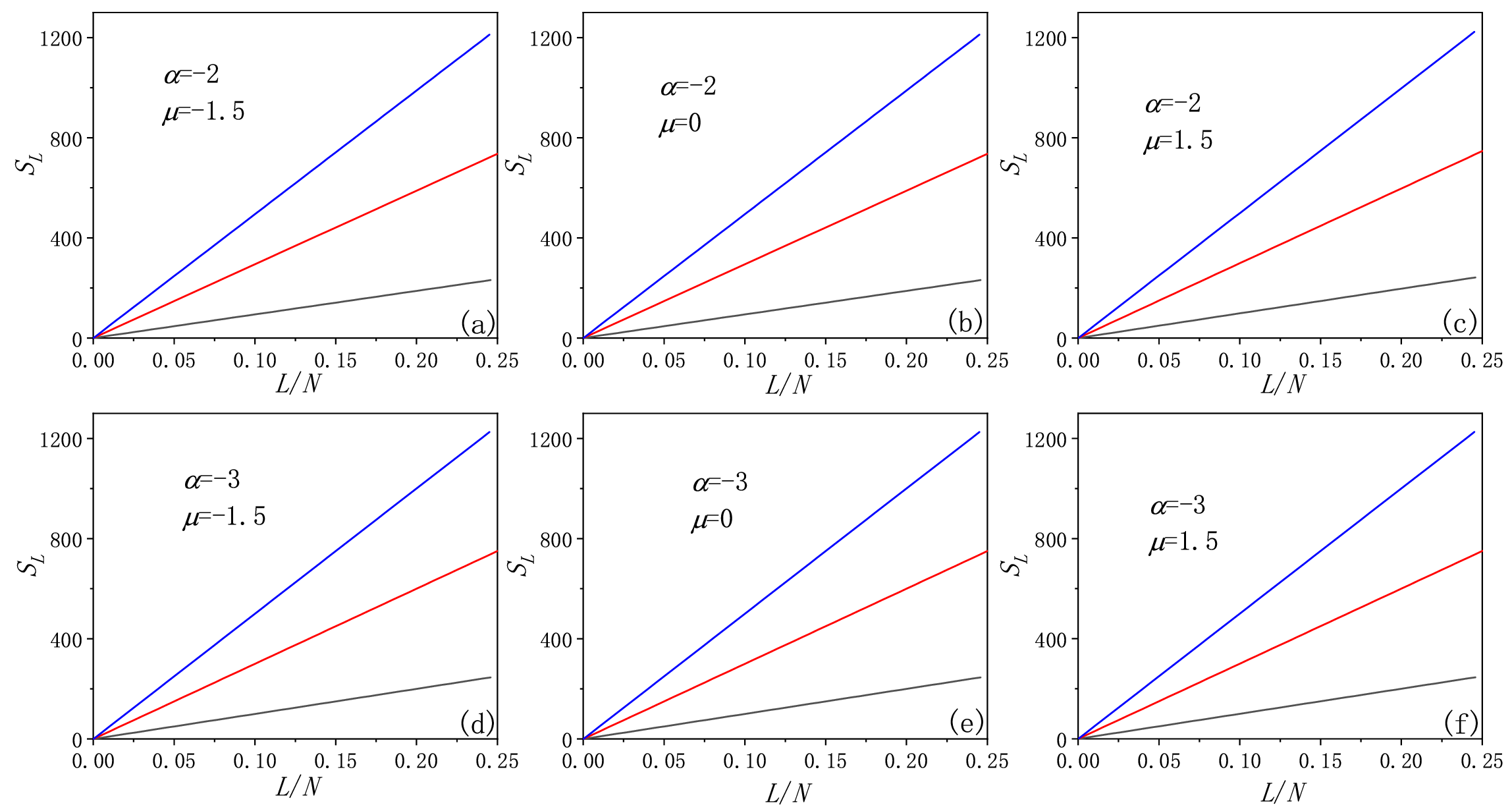}
 \caption{\label{s5} (Color online) The von Neumann entropy $S_L$ as functions of the $L/N$ for $\alpha\le-2$.
 The black, red, and blue lines represent to $N=1001$, $3001$, and $5001$, respectively.
 }
\end{figure}

\par Similar to that in the case of $-2<\alpha<0$ in Fig. \ref{s2} and Fig. \ref{s3}(a)-(d), the von Neumann entropy $S_L$ as functions of the relative size of subchain $L/N$ are shown in Fig. \ref{s5} for $\alpha\le-2$.
From the figures it is found that $S_L$ is proportional to $L$.
It means that the relationship between $S_L$ and $L$ is also fitted by $S(L)=A(L-L_0)^{\beta}$ with $\beta\approx1$.

\bibliographystyle{apsrev4-1}
\bibliography{test2}